\documentclass[preprint,12pt]{article}
\usepackage{latexsym,epsfig,amssymb,amsmath,subfigure,color,pictex,fullpage,amsfonts}
\usepackage[english]{babel}
\usepackage[latin1]{inputenc}
\usepackage{times}
\usepackage[T1]{fontenc}
\usepackage{graphicx}
\usepackage{epsfig}
\usepackage{graphics}
\usepackage{amssymb}
\usepackage{amsmath}
\usepackage{multicol}
\usepackage{pst-grad}
\usepackage{pst-plot}
\usepackage{pstricks}
\usepackage{epsfig}
\usepackage{setspace}
\usepackage{xcolor}
\usepackage{float}
\usepackage{epstopdf}
\usepackage{authblk}
\usepackage{booktabs}
\usepackage{enumerate}
\usepackage[normalem]{ulem}

\usepackage{xcolor,cancel}

\renewcommand{\mathit}[1]{\mbox{\em #1}}

\renewcommand{\mathit}[1]{\mbox{\em #1}}

\newcommand{\ds}{\displaystyle}
\newcommand{\be}{\begin{eqnarray}}
\newcommand{\ee}{\end{eqnarray}}

\newcommand{\non} \nonumber
\newcommand{\dd} \partial

\newcommand{\ie}{{\it i.e.}\ }

\newcommand{\on}{\quad \mbox{on} \quad}

\newcommand{\for}{\quad \mbox{for} \quad}

\definecolor{pink}{rgb}{1,0.5,0.5}
\definecolor{darkred}{rgb}{0.8, 0.0, 0.0}
\definecolor{darkgreen}{rgb}{0.0, 0.6, 0.0}
\definecolor{darkblue}{rgb}{0.0, 0.0, 0.8}
\definecolor{grey}{rgb}{0.7 0.7 0.7}

\newcommand{\bx}{\mathbf{x}}
\newcommand{\bX}{\mathbf{X}}

\newcommand{\po}{p^{(0)}}
\newcommand{\pone}{p^{(1)}}
\newcommand{\ptwo}{p^{(2)}}

\newcommand{\uo}{u^{(0)}}
\newcommand{\uone}{u^{(1)}}

\newcommand{\yo}{y^{(0)}}
\newcommand{\yone}{y^{(1)}}

\newcommand{\Uo}{U^{(0)}}
\newcommand{\Uone}{U^{(1)}}
\newcommand{\Tho}{\Theta^{(0)}}

\newcommand{\sigmao}{\sigma^{(0)}}
\newcommand{\sigmaone}{\sigma^{(1)}}
\newcommand{\sigmaeff}{\sigma^{\text{eff}}}

\newcommand{\peff}{p^{\text{eff}}}

\newcommand{\ueff}{u^{\text{eff}}}
\newcommand{\epsiloneff}{\epsilon^{\text{eff}}}

\newcommand{\Seff}{S^{\text{eff}}}
\newcommand{\Eeff}{E^{\text{eff}}}
\newcommand{\seff}{s^{\text{eff}}}
\newcommand{\eeff}{e^{\text{eff}}}

\newcommand{\epsilono}{\epsilon^{(0)}}
\newcommand{\So}{S^{(0)}}
\newcommand{\Eo}{E^{(0)}}
\newcommand{\so}{s^{(0)}}
\newcommand{\eo}{e^{(0)}}

\newcommand{\oo}{^{(0)}}

\newcommand{\ginc}{\Gamma_{\text{inc}}}

\newcommand{\PP}{{\mathcal P}}
\newcommand{\BB}{{\mathcal B}}

\title{Multiple scales homogenisation of a porous viscoelastic material with rigid inclusions: application to lithium-ion battery electrodes}

\author[1,3,6]{J. M. Foster}
\author[1,2]{A. F. Galvis}
\author[4]{B. Protas}
\author[3,5]{S. J. Chapman}
\affil[1]{School of Mathematics and Physics, University of Portsmouth, Portsmouth, PO1 2UP, UK}
\affil[2]{School of Electrical and Mechanical Engineering, University of Portsmouth, Portsmouth, PO1 3DJ, UK}
\affil[3]{The Faraday Institution, Quad One, Harwell Science and Innovation Campus, Didcot, OX11 0RA, UK}
\affil[4]{Department of Mathematics and Statistics, McMaster University, Hamilton, Ontario, Canada}
\affil[5]{Mathematical Institute, University of Oxford, OX2 6GG, UK}
\affil[6]{email: jamie.michael.foster@gmail.com}

\begin{document}
	
	
\maketitle


	This paper explores the mechanical behaviour of the composite
	materials {used} in modern lithium-ion battery electrodes. These
	contain relatively high modulus active particle inclusions within a
	two-component matrix of liquid electrolyte which penetrates the pore
	space within a viscoelastic polymer binder. Deformations are driven by
	a combination of (i) swelling/contraction of the electrode particles
	in response to lithium insertion/extraction, (ii) swelling of the
	binder as it absorbs electrolyte, (iii) external loading and (iv) flow
	of the electrolyte within the pores. We derive the macroscale response
	of the composite using systematic multiple scales homomgenisation by
	exploiting the disparity in lengthscales associated with the size of
	an electrode particle and the electrode as a whole. The resulting
	effective model accurately replicates the behaviour of the original
	model (as is demonstrated by a series of relevant case studies) but,
	crucially, is markedly {simpler and hence} cheaper to solve. This
	is significant practical value because it facilitates low-cost,
	realistic computations of the mechanical states of battery electrodes,
	thereby allowing model-assisted development of battery designs that
	are better able to withstand the mechanical abuse encountered in
	practice and ultimately paving the way for longer-lasting batteries.
	
	
		

\section{Introduction}
{Lithium-ion batteries (LIBs) are already ubiquitous in many rechargeable energy storage applications, including consumer electronics, off-grid storage and increasingly in the use of electric vehicles. They provide a high energy- and power-density, a high cell voltage and effectively maintain their charge when not in use \cite{Kan06}. Despite this, there is a significant global impetus to make further improvements in safety, length of usable lifetimes and in the ability to facilitate high (dis)charging rates. It is expected that LIBs will be a key technology in realising a truly sustainable and low-carbon economy in the coming decades, and so these improvements need to be developed promptly \cite{Zha19}.}

{LIBs {comprise} a number of connected cells. When a pouch format
  is used, planar cells are stacked one on top of the other, whereas
  in a cylindrical roll format (e.g., the popular 18650) cells are
  wound in a spiral around a central rod. Regardless of the format,
  each individual cell {comprises} two porous electrodes (one
  anode and one cathode) which are electrically insulated from one
  another by a porous separator diaphragm and this assembly is then
  sandwiched between two metallic current collectors. The pore space
  is filled with a liquid electrolyte and this provides a pathway for
  Li-ions to move from one electrode to the other. During discharge,
  Li-ions move from the anode to the cathode and this migration of
  positive charge is offset by the motion of electrons which also
  travel from the anode to cathode. Since the separator is
  electrically insulating the electrons migrate via the current
  collectors and external circuit and it is this electrical current
  that we utilise to power devices. The {charging} process works
  similarly, but it requires the application of an external voltage
  which provides the requisite incentive for the Li-ions and electrons
  to move from the cathode to the anode \cite{Owe97}.}  

{In modern LIBs both electrodes are multiphase and contain (i) active
  material particles which act as a reservoir for Li storage, (ii) the
  liquid electrolyte, and (iii) a polymer binder. {The latter is often
    doped with carbon black nanoparticles which provides a boost to
    the electronic conductivity so that the binder can transport
    electrons without dropping much voltage. The polymer binder also
    provides the electrode with structural integrity and helps to
    maintain the contact between the internal particles {in
      addition to} connecting the electrode with the current
    collector.} This composite electrode is multiscale in
  nature. There are several distinct lengthscales relevant to each
  electrode, namely, (a) the electrode extent in the direction
  parallel to the current collectors which is $O(10^{-2}\:\text{m})$,
  (b) the electrode thickness in the direction perpendicular to
  current collectors which is typically $O(10^{-4}\:\text{m})$, (c)
  the radius of an individual particle of active material which is
  often $O(10^{-6}\:\text{m})$, and (d) typical pores within the
  binder which are of size $O(10^{-7}\:\text{m})$. The disparity in
  these lengthscales plays a central role in this work.} 

Several factors contribute towards the undesirable loss of efficacy of a LIB throughout its lifetime \cite{Vet05}. Chemical forms of degradation can begin occurring as early as the very first cycle, known as the formation cycle, during which a solid electrolyte interphase (SEI) layer forms on the surface of the active material particles \cite{Eta11,Saf09}. This results in a loss of mobile Li which therefore limits the amount of useful charge that can be stored/extracted. Other chemical degradation mechanisms continue throughout the {cell's} service and can include lithium plating and dendrite formation, dissolution and corrosion of the current collectors, and disordering within the active materials \cite{Bir17}. Alongside these chemical forms of degradation there are also mechanical modes of damage; excellent reviews of these can be found in \cite{McD16,Zha19}. The onset of this form of aging can begin before the cell has even been assembled. During manufacture, once an electrode has been made, it is subjected to calendering where the electrode is squeezed between two rollers at high pressure in an effort to remove some of the superfluous pore space, thereby increasing energy density \cite{Len17,Kwa18,Kes15,Wan04,Gim19,Gim20}. However, there is strong evidence that this calendering can cause active particles to fracture or become fatigued \cite{Ant15,Zhe12,Has13}. When the cell is subsequently assembled and exposed to electrolyte for the first time, the polymers that are used to bind the electrodes swell, inducing further internal stresses \cite{Che06,Cho14,Mag10}. More loading occurs when the cell is in use and the insertion/removal of Li from the active material particles causes volumetric swelling/shrinkage \cite{Kim04,Xia11,Ebn13}. Together these factors have been shown to cause a variety of forms of damage including delamination of the electrode from the current collector, disconnection of the particles from the surrounding binder \cite{Liu12,Kov11,Shi02,Che13}, cracking of the particles themselves, and in extreme cases (sometimes due to a significant external impact) internal short circuits on puncturing of the separator \cite{San09,Cai11}.

{Many previous modelling efforts have aimed to understand mechanics at one of the relevant lengthscales. For example, authors have presented models for the stresses that are induced within individual electrode particles, and these models often aim to predict intra-particle crack formation and propagation, see e.g. \cite{Chr06,Chr06b,Zha07,Kli16}. Other work, at the scale of electrodes, has examined how stresses within the particles influences electrochemical characteristics and hence cell performance \cite{Ai20,Rie16}. {In \cite{Timms2022} the stresses due to thermal and chemical expansion are examined at the level of a battery, specifically a spirally-wound cylindrical roll. Finally, another}  fertile area is mechanics at the level of entire packs where a common aim is to make use of stress/strain measurements that can be taken in-situ to estimate states of charge or health \cite{Oh16,Dai17}. Whilst all these investigations are  highly valuable, a complete theory of LIB mechanics can only be achieved if {phenomena occurring at} these vastly different lengthscales can be reconciled. Some authors have made progress here by adopting computational approaches in which models are solved on representative sections of the battery, see for example \cite{Wu14}, but, these methods become prohibitively time-consuming to solve as the size and complexity of the sample geometry increases. }  

{An alternative approach, and the one we shall adopt here, is to make
  use of homogenisation methods which provide an efficient way to
  model the behaviour of heterogeneous materials. There are a variety
  of approaches to carrying out this homogenisation and an excellent
  review of the different strategies and their use in studying the
  mechanics of solid composites is given in \cite{Kan09}. In this
  study we use the asymptotic method of multiple scales. The
  overarching strategy is to begin by posing model equations on the
  microstructured geometry---all the physics of the problem is written
  for the constituents at the microscale. In the present case this
  includes swelling of the binder and active material particles
  (caused by (de-)lithiation), viscoelastic deformation in the binder
  (caused by absorption of the electrolyte), Darcy flow of the
  electrolyte within the binder's pores and different damage states on
  the interfaces between the particles and binder. So-called small-
  and large-scale spatial variables are then introduced and are
  assumed to be independent, and it is further assumed that the
  microstructure is locally periodic. In this case the large-scale
  spatial variable is one that varies on the scale of the whole
  electrode whereas the small-scale one varies on the scale of
  individual electrode particles. A perturbation problem results from
  these assumptions which gives rise to conservation and force balance
  equations that govern the equivalent homogenised medium. Solutions
  on the small-scale are resolved on a representative volume element
  (RVE) and these establish constitutive equations for the homogenised
  medium. In our case the result is a mechanical model that can
  simultaneously capture the lengthscales of individual electrode
  particles up to those of the whole electrode, but, crucially, it
  remain relatively {simple and therefore} cheap to solve.} 

{The problem that we consider here is closely related to a pair of
  prior studies \cite{Fos17,Fos17b} which also consider how to relate
  the deformations that can occur on the lengthscale of individual
  active material particles to those on the scale of the whole
  electrode. Like the present study, the active material particles and
  binder phases were resolved. The former material is markedly stiffer
  than the latter and so the particles were assumed to be rigid
  bodies, whilst the binder was assumed to be viscoelastic
  \cite{Che04,Wu14}. Similar assumptions are made here but we also
  introduce two important novelties. First, the homogenisation
  performed here is systematic, whereas in
  \cite{Fos17,Fos17b} this was done on an ad-hoc basis. This allows us
  to explicitly interrogate how changes to microscale equations
  influence the properties on the macroscale scale and opens the
  possibility to identify material/geometrical properties on the
  microscale that give rise to desirable characteristic on larger
  scales. Second, we include coupling between the liquid electrolyte
  and solid materials and, as we will demonstrate, the loading of the
  solid matrix when the electrolyte is forced to flow can become
  significant when electrodes are subjected to sudden impact.} 

This work also fits within the broader field of the mechanics of
composites. {The mathematical aspects of this field are surveyed
  in the classic monographs
  \cite{Cherkaev2000,Milton2002,Torquato2002}.}  Notable related
studies include \cite{Bur81} {which derives the Biot poroeasticty
  model} from the equations describing an elastic medium with Stokes
flow in the pores using multiples scales homogenisation.  Several
studies have considered the overall response of a medium comprised of
spherical particles embedded within an elastic {matrix}, see
e.g.~\cite{Ogd74}. Others {looked} at the effects of
incorporating damage, either within the bulk of one of the constituent
phases \cite{Dvo00,Dvo01,Voy95}, or on the interface between the
different materials \cite{Fis99,Bul99,Fis01}. Some studies
{compared} the multiscale models that result from different
homogenisation strategies \cite{Seg02,Dav13,Fey99}. Finally,
{there is also research that} is highly application driven and
work has been done to understand concretes and cement paste
\cite{Hai08,Con16}, metal-matrix composites \cite{Hu98} and
{certain related problems} in bio-mechanics \cite{Par06,Mul06}.

The structure of the remainder of this work is as follows. In
\S\ref{sec2} we formulate the governing equations on the
microstructured geometry {and}, in \S\ref{sec3},
nondimensionalise the equations and provide estimates on the
dimensionless parameters that result. {Next,} in \S\ref{sec:ms}
we carry out the multiple scales analysis, using a combination of
analytical and numerical solutions to the requisite problem on an 
RVE. In \S\ref{results} we find some solutions to the
effective models in {certain} scenarios relevant to realistic LIB
assembly and useage, and demonstrate that the effective model
faithfully reproduces the original. Finally, in \S\ref{sec7} we draw
our conclusions. 

\section{\label{sec2}Model formulation}

\begin{figure}
\centering
\includegraphics[width=\textwidth]{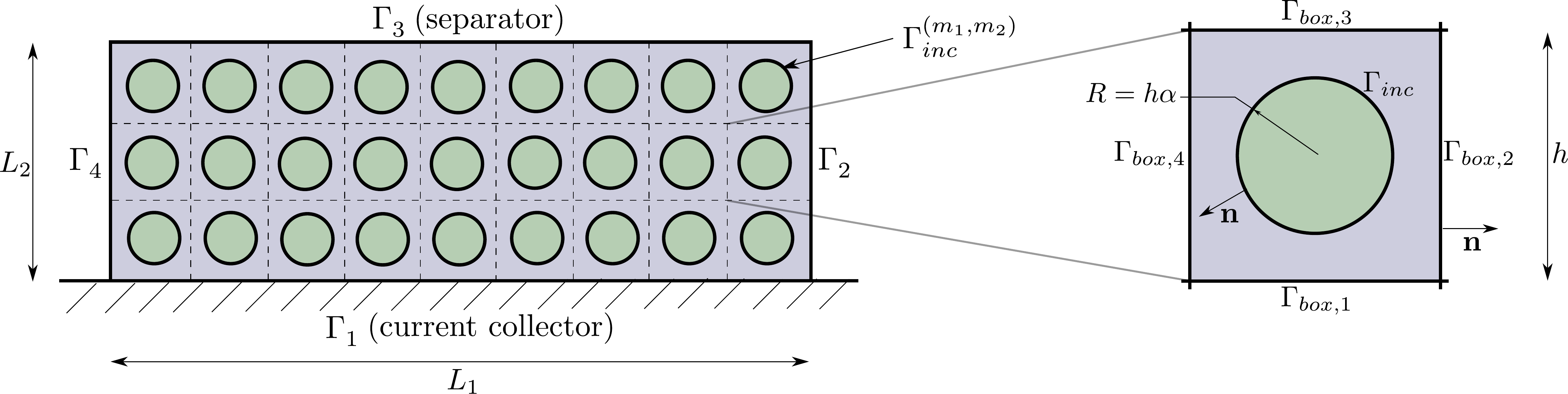}
\caption{Left: A sketch of the geometry. Right: An expanded view of a
  representative volume element. The radius of the particle is given
  by $R = h\alpha$ where $\alpha \in (0,1/2)$ {and $h$ is the
    size of the volume element}. The boundary segments $\Gamma_{1-4}$
  correspond to the external edges of the electrode and
  $\ginc^{(m_1,m_2)}$ the interfaces between the particles and porous
  binder with the indices {$m_1,m_2 \in \mathbb{Z}$} labelling
  the different particles.}
\label{fig:figure1}
\end{figure}

An idealised electrode is shown in Fig.~\ref{fig:figure1}. For
simplicity we will work in two spatial dimensions but we emphasize
that the approach readily extends to higher dimensions. The electrode
particles will be assumed to be cylinders of equal radius whose centres
lie on a regular square lattice.

First we shall formulate the equations to be solved in the porous
binder, then we state boundary conditions that couple the porous
binder to the electrode particles embedded within it. The particles
shall be assumed to be rigid bodies, {such that} the model is
closed by supplying force and torque balances to determine their rigid
body translations and rotations. With the model specified, we shall
nondimensionalise {it} and provide estimates for the
dimensionless parameters. Finally, we summarise the dimensionless model
before proceeding with the multiple scales homogenisation in
\S\ref{sec:ms}. {Hereafter, we will use Einstein's convention
  with repeated indices implying summation; the indices will take
  values $i,j \in \{1,2\}$.}

\subsection{Equations for the porous binder and electrolyte} 
The force balance equations and definition of the (small/infinitesimal) strain are
\be
\label{gov1} \frac{\partial }{\partial x_j} \left( \sigma_{ij} {- \delta_{ij} p} \right) =0, \quad \epsilon_{ij} = \frac{1}{2} \left( \frac{\partial u_i}{\partial x_j} + \frac{\dd u_j}{\dd x_i} \right),
\ee
where $\sigma_{ij}$ and $\epsilon_{ij}$ are the stress and strain within the solid skeleton (binder) respectively, $p$ is the fluid (electrolyte) pressure, and $u_i$ is the deformation in the solid skeleton. The former equation in (\ref{gov1}) is arrived at by considering a RVE that is small by comparison to the lengthscale of an inclusion, but large by comparison to the scale of a typical pore within the porous skeleton material \cite{biot,bear,wang}. Definitions of the volumetric and deviatoric parts of the stress ($S$ and $s_{ij}$ respectively) and strain ($E$ and $e_{ij}$ respectively) are
\be
\label{gov2} S = \frac{1}{d} \sigma_{kk}, \quad E = \frac{1}{d} \epsilon_{kk},\\
\label{gov3} s_{ij} = \sigma_{ij} - \delta_{ij} S,\quad e_{ij} = \epsilon_{ij} - \delta_{ij} E.
\ee
Here, $\delta_{ij}$ is the Kronecker $\delta$, and $d$ is the number
of spatial dimensions. Henceforth, we shall set $d=2$, but shall signpost
in various places where the results depend upon the number of spatial
dimensions. The binder is assumed to be a standard linear
viscoelastic, and as such its constitutive relation is given
by\footnote{An equivalent statement of the constitutive laws can be
  made via hereditary integrals, namely $S=E(0)K(t) + \int_0^t
  K(t-\tau) \dot{E}(\tau) d\tau$ where $K(t)  = K_1 + (K_2-K_1)
  \exp(-t/K_\tau)$ and similarly for the deviatoric components. Thus,
  $K_1$ and $K_2$ are the long-term and instantaneous moduli
  respectively and $K_\tau$ is the relaxation timescale.} 
\be
\label{gov4} {G_{\tau} \dot{s}_{ij} + s_{ij} = G_2 G_\tau \dot{e}_{ij} + G_1 e_{ij},} \quad \label{govend} {K_{\tau} \dot{S} + S = K_2 K_\tau \left( \dot{E} - \dot{\beta} \right) + K_1 (E-\beta).}
\ee
Here, $\beta$ is an eigenstrain induced by the swelling of the polymer binder as it absorbs the liquid (electrolyte) immediately after cell construction. The motion of the fluid in the pore space is governed by Darcy's law, \ie
{
  \begin{equation}
  \phi_f \left( w_i - \frac{\dd u_i}{\dd t}\right) = {-\frac{k}{\mu}} \frac{\dd p}{\dd x_i} \label{Darcy}
\end{equation}
}
where $\phi_f$ is the volume fraction of the liquid, $w_i$ is the
fluid velocity, $k$ is the permeability of the solid skeleton and
$\mu$ is the liquid viscosity. We assume that the liquid in
incompressible. This, combined with our assumptions on small strains (
see (\ref{gov1})) leads to the following {equations for 
  the conservation} of mass (and hence volume): 
{
  \begin{equation}
  \frac{\dd \phi_f}{\dd t} + \frac{\dd}{\dd x_i} ( w_i \phi_f ) =0,
  \qquad \frac{\dd}{\dd t} (1-\phi_f) + \frac{\dd}{\dd x_i} \left( \frac{\dd u_i}{\dd t} (1- \phi_f) \right) =0.\label{conservation}
\end{equation}
}
{Using} the standard procedure of eliminating the fluid velocity, $w_i$, as well as the volume fraction $\phi_f$  between {(\ref{Darcy}) and (\ref{conservation})} leads to \cite{Fow97}
\be \label{gov5}
{\frac{\partial}{\partial t} \left( \frac{\partial u_i}{\partial x_i} \right) = \frac{k}{\mu} {\frac{\partial^2 p}{\partial x_j\partial x_j }}}.
\ee
Equations {(\ref{gov1})--(\ref{gov4}), (\ref{gov5})} {represent fifteen scalar relations} (five  in (\ref{gov1}), two in (\ref{gov2}), four in (\ref{gov3}), three in (\ref{gov4}) and one in (\ref{gov5})) in fifteen unknowns (two displacements $u_i$, three stresses $\sigma_{ij}$, three strains $\epsilon_{ij}$, a volumetric stress $S$, a volumetric strain $E$, two deviatoric stresses $s_{ij}$, two deviatoric strains $e_{ij}$, and a pressure $p$).

\subsection{\label{BCss}Boundary conditions}
We model the electrode particles as rigid bodies {which is} justified on the basis that they are markedly stiffer than the binder. Therefore, boundary conditions on the electrode particle centred at
$(x_1,x_2)=(h m_1,h m_2)$ with boundary $\Gamma^{(m_1,m_2)}$ (see Fig.~\ref{fig:figure1}) are 
\be \label{BC} 
u_i = (x_i - h m_i) g(t) + U^{(m_1,m_2)}_i(t) +\Theta^{(m_1,m_2)}(t) \varepsilon_{ij3} (x_j - h m_j) \quad \on \Gamma_{inc}^{(m_1,m_2)}, \\ \frac{\dd p}{\dd x_i} n_i = 0 \on
\Gamma_{inc}^{(m_1,m_2)}.  \label{BC2}
\ee 
The former condition states that the binder skeleton is bonded to the electrode particle such that there is continuity of displacement on the interface between the two materials \cite{Cha90} and the latter requires that the fluid and {solid} velocities {be} equal, so that there is no flow into/out of the electrode particle. Here $g(t)$ characterises the expansion/shrinkage of the electrode particle in response to (de)lithiation. The condition (\ref{BC}) contains an implicit assumption that the particles change their volume in a  {radially-symmetric manner}. The quantities $U_i^{(m_1,m_2)}(t)$ are the translations and $\Theta^{(m_1,m_2)}(t)$ is a {clockwise rotation} of the particles, and $\varepsilon_{ijk}$ is the Levi-Civita symbol. The translations $U_i^{(m_1,m_2)}(t)$ and rotation $\Theta^{(m_1,m_2)}$ {are a priori unknown and} must be determined as part of the solution to the problem using the following force and torque balance constraints \cite{Cha15}
\be
\label{constraint} \int_{\Gamma_{inc}^{(m_1,m_2)}} \left( \sigma_{ij}  {- \delta_{ij} p } \right) n_j \, d\Gamma = 0 \qquad \mbox{and} \qquad \int_{\Gamma_{inc}^{(m_1,m_2)}} \varepsilon_{ik3} (x_k-h m_k) \sigma_{ij} n_j \, d\Gamma = 0.
\ee
We require boundary conditions on the outer edges of the electrode as well as initial conditions to complete the model, however, the multiple scales analysis that is the main focus of this work is independent of the form of these conditions. In the interests of generality we shall proceed, for the time being, without specifying these conditions. Later, in \S\ref{results}, we shall return to this issue and present the remaining boundary and initial conditions for some representative scenarios.


\subsection{\label{sec3}Nondimensionalisation}
We write the problem in dimensionless form by introducing the following scalings
\begin{align} \label{scalings}
x_i &= L_2 x_i^* , 					& t &= \tau t^*,	& \Theta^{(m_1,m_2)} &= \frac{{\mathcal U}_0}{L_2} \Theta^{*(m_1,m_2)}, 				\\	
u_i &= {\mathcal U}_0 u_i^*,				& U_i^{(m_1,m_2)} &= {\mathcal U}_0 U_i^{*(m_1,m_2)},				&  p &= \frac{\mu L_2 {\mathcal U}_0}{k \tau} p^*,\\
\epsilon_{ij} &= \frac{{\mathcal U}_0}{L_2} \epsilon_{ij}^*,		& e_{ij} &= \frac{{\mathcal U}_0}{L_2} e_{ij}^*,		& E &= \frac{{\mathcal U}_0}{L_2} E^*,\\
\sigma_{ij} &= \frac{Y {\mathcal U}_0}{L_2} \sigma_{ij}^*,					& s_{ij} &= \frac{Y {\mathcal U}_0}{L_2} s_{ij}^*,					& S &=  \frac{Y {\mathcal U}_0}{L_2} S^*,\\ 	
\beta &= {\mathcal B}_0 \beta^*,						& g &={\mathcal G}_0 g^* .\label{sc}		
\end{align}
{where} a star indicates a dimensionless quantity. Here, $L_2$ is the thickness of the electrode, $Y$ is the typical modulus of the binder, and ${\mathcal U}_0$ is the typical deformation. The timescale of interest, $\tau$, may be one of several different {quantities}, including: the timescale of an impact ($\sim$ 1 second), the timescale of electrochemical cycling ($\sim$ 1 hour), the timescales associated with the swelling of the binder after being immersed in electrolyte ($\sim$ 10 hours), or the timescale associated with calender aging ($\sim$ 1 year). Scaling in this way introduces the following dimensionless parameters which characterise the problem
\begin{align}
\Delta^* &= \frac{h}{L_2},				& {\mathcal P}^* &= \frac{\mu L_2^2}{k \tau Y},  & {\mathcal B}^* &= \frac{{\mathcal B}_0 L_2}{{\mathcal U}_0}  & {\mathcal G}^* &= \frac{{\mathcal G}_0 L_2}{{\mathcal U}_0},\\
G_\tau^* &= \frac{G_\tau}{\tau},		& K_\tau^* &=\frac{K_\tau}{\tau}, &G_i^* &= \frac{G_i}{Y},				& K_i^* &= \frac{K_i}{Y},
\end{align}
{where} the index $i=1,2$ in the final two equations corresponds
to the different moduli rather than a spatial dimension. The only
dimensionless parameter whose meaning is not self-evident from its
definition is {${\mathcal P^*}$}; {this may be interpreted
as the ratio of the pressure in the electrolyte generated by the
motion of the binder to the elastic stresses in the binder (which is
how it appears in the equations below), but it is perhaps better
interpreted as the ratio of the relaxation time for pressure gradients
to the timescale of interest (it would appear on the left-hand side of
(\ref{gov5nd}) after rescaling $p^*$ with ${\mathcal P^*}$).}

\subsection{Parameter values}

{Our analysis  will be conducted in the limit $\Delta \ll 1$, corresponding to electrode particle radii much smaller than electrode thickness.}
Typically electrode particle radii are $O(1-10)\mu$m whilst electrode thickness are $O(100)\mu$m yielding an estimate of \begin{equation}
\Delta = O(10^{-1} - 10^{-2}).
\end{equation}
Whilst this parameter is small, one might be concerned that the contrast in lengthscales is not sufficiently large to yield a homogenised model that is very accurate. As we will demonstrate later, in \S\ref{results}, the convergence between the homogenised and full models appears rather rapid: discrepancies between the homogenised model and the full model are $<1$\% even when $\Delta = O(10^{-1})$.

We justify our assumption that the particles may be modelled as rigid bodies by noting that the moduli of the most common binder, namely polyvinylidene fluoride (PVDF), is $O(1-10)$MPa \cite{Che04,Wu14} whilst that of NMC, as well as other common electrode materials, is on the order of $O(10-100)$GPa. As such, it is reasonable to expect that the porous binder undergoes very significant deformations before the particles deform appreciably. An exception might be scenarios where particles come into contact with one another and external loads are high; however, particle-particle contact is not considered {here}  and will instead be the subject of future work.

All other dimensionless parameters will be taken to be of $O(1)$ so that the ensuing analysis is relevant in the most general case in which deformations driven by the swelling of the binder, the cycling of the electrode particles and both bulk and shear stresses are all significant (the distinguished limit). We note that any of these can be taken to be negligible after the homogenisation procedure.

{We note that in \cite{Fos17} the limit ${{\mathcal
      P}^*} \to 0$ was taken. If the timescale of interest is
  relatively long, as would be the case for long-term aging or even
  cell cycling, then studying the solution in the limit that
  ${{\mathcal P}^*} \to 0$ is likely to be appropriate.
{For example, at a charge/discharge rate of $C/10$ with an active
  material expansion/contraction of $10\%$ we estimate ${\cal P}^*
  =O(10^{-4})$. }
  In this situation, pressure gradients are instantaneously relaxed by
  fluid flowing through the pore space, and consequently the equations
  governing the stresses in the solids decouple from those in the
  liquid. In what follows, {however}, we will study the distinguished
  limit in which ${{\mathcal P}^*}=O(1)$ so that there is
  coupling between the solid and fluid. By doing so, our analysis is
  applicable {also} to situations in which the deformations are
  rapid, as might be the case if the electrode is subjected to an
  sudden impact or when the electrode is calendered during the
  manufacturing process. {For example, we estimate that  a strain of
  $10\%$ applied over a timescale of $1s$ (in a crash, say) would give ${\mathcal P}^*=O(1)$.}


\subsection{The dimensionless model}
On applying the scalings (\ref{scalings})--(\ref{sc}) to (\ref{gov1})--(\ref{constraint}) we arrive at the dimensionless system
\be
\label{gov1nd} \frac{\partial }{\partial x_j^*} \left( \sigma_{ij}^* {- {\mathcal P}^* \delta_{ij} p^*} \right) & = & 0, \\
\epsilon_{ij}^* & = & \frac{1}{2} \left( \frac{\partial u_i^*}{\partial x_j^*} + \frac{\dd u_j^*}{\dd x_i^*} \right),\\
\label{gov2nd} S^* & = & \frac{1}{2} \sigma_{kk}^*, \\
E^* & = & \frac{1}{2} \epsilon_{kk}^*,\\
\label{gov3nd} s_{ij}^* & = & \sigma_{ij}^* - \delta_{ij} S^*, \\
e_{ij}^* & = & \epsilon_{ij}^* - \delta_{ij} E^*,\\
\label{benny} G_{\tau}^* \dot{s}_{ij}^* + s_{ij}^* & = & G_2^* G_\tau^* \dot{e}_{ij}^* + G_1^* e_{ij}^*,\\ 
\label{noemmys} K_{\tau}^* \dot{S}^* + S^* & = & K_2^* K_\tau^* \left( \dot{E}^* - {\mathcal B}^* \dot{\beta}^* \right) + K_1^* (E^*-{\mathcal B}^* \beta^*),\\
\label{gov5nd} \frac{\partial}{\partial t^*} \left( \frac{\partial u_i^*}{\partial x_i^*} \right) & = & \frac{\partial^2 p^*}{\partial x_i^{*} \partial x_i^{*}},
\ee
with boundary conditions
\be
\label{cheesy} u_i^*  = (x_i^* -  \Delta^* m_i) {\mathcal G}^* g^*(t) + U^{*(m_1,m_2)}_i(t^*) +\Theta^{*(m_1,m_2)}(t^*) \varepsilon_{ij3}  (x_j^* - \Delta^* m_j)  \on \Gamma_{inc}^{*(m1,m2)}, \\
\label{feta} \frac{\dd p^*}{\dd x_i^*} n_i =  0  \on
\Gamma_{inc}^{*(m1,m2)},  
\ee
and we reiterate that additional initial conditions as well as boundary conditions, which are yet to be specified, are required at the outer edges of the electrode {$\Gamma_1$, $\Gamma_2$, $\Gamma_3$  and $\Gamma_4$.} The displacement and rotation of each inclusion are determined by the force and torque constraints
\be
\label{constraintnd} \int_{\Gamma_{inc}^{*(m_1,m_2)}} \left( \sigma_{ij}^*
 {- {\mathcal P}^* \delta_{ij} p^* } \right) {n}_j \, d\Gamma^* = 0 \quad \mbox{and} \quad
\int_{\Gamma_{inc}^{*(m_1,m_2)}} \varepsilon_{ik3} (x_k^*-\Delta^* m_k) \sigma_{ij}^* n_j
    \, d\Gamma^* = 0.
\ee
{To simplify notation, hereafter we will drop the star from dimensionless quantities.}


\section{\label{sec:ms}Multiple scales homogenisation}
{We now proceed with the homogenisation of {system} (\ref{gov1nd})--(\ref{constraintnd}) using the asymptotic method of multiple scales in the limit $\Delta \ll 1$.
We introduce microscopic spatial variables by setting
\be
X_i = \frac{x_i}{\Delta}.
\ee
}
\begin{figure}
\centering
\includegraphics[scale=0.2]{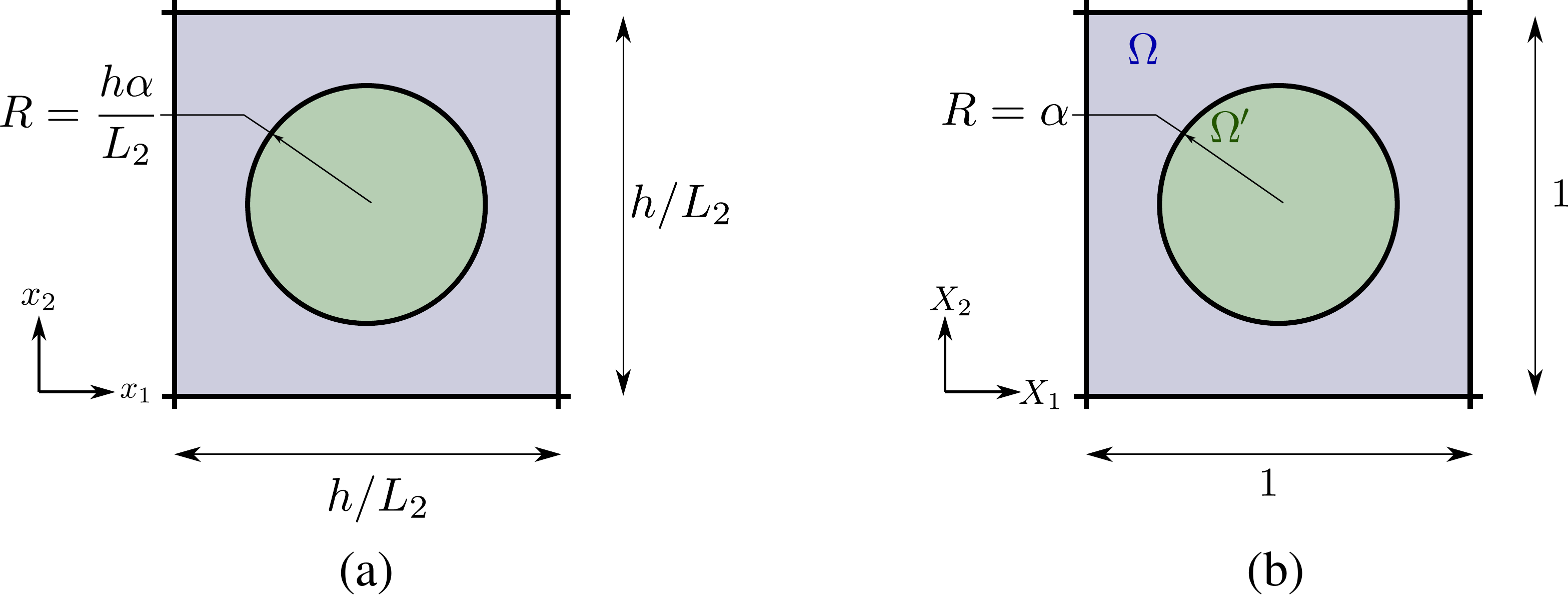}
\caption{A sketch of the RVE in (a) the dimensionless macroscale coordinates and (b) the dimensionless microscale coordinates.}
\label{RVEs}
\end{figure}
{As is usual in the multiple-scales approach, we assume that $x_i$ and $X_i$ are independent, with derivatives transforming according to the chain rule
\be \label{snowdog}
 \frac{\dd}{\dd x_j} \rightarrow \frac{\dd}{\dd x_j} + \frac{1}{\Delta} \frac{\dd}{\dd X_j}.
\ee
We remove the freedom that this introduces by imposing that all variables are exactly periodic in $X_i$ with period 1, with deviations from periodicity taken up by modulation in the slow variable $x_i$.
Some thought is needed to write the displacements and rotations of the inclusions, $U_i$ and $\Theta_i$ in multiple-scales form. These are functions of time only, but may vary from inclusion to inclusion, so will become functions of space after the homogenisation procedure.
We impose that they are functions of time only by writing
\[ \frac{\dd U_i }{ \dd x_j}=\frac{\dd \Theta}{\dd x_j} = 0,\]
which is now in a form suitable for transforming to multiple scales, becoming
\be
\label{new} \frac{1}{\Delta} \frac{\partial U_i}{\partial X_j} + \frac{\partial U_i}{\partial x_j} = 0, \qquad  \frac{1}{\Delta} \frac{\partial \Theta}{\partial X_j} + \frac{\partial \Theta}{\partial x_j} = 0.
\ee
In principle we could do the same with the swelling $g(t)$, but since in our model this function is imposed rather than to be determined we do not need to.
}
The governing PDEs (\ref{gov1nd})--(\ref{gov5nd}) become
\be
\label{ms1} \frac{1}{\Delta} \frac{\partial}{\dd X_j} \left( \sigma_{ij} - \PP \delta_{ij} p \right) & = & -\frac{\dd}{\dd x_j} \left( \sigma_{ij} - \PP \delta_{ij}  p \right), \\
\label{ms1b}
\epsilon_{ij} & = & \frac{1}{2\Delta} \left( \frac{\partial u_i}{\partial X_j} + \frac{\dd u_j}{\dd X_i} \right) + \frac{1}{2} \left( \frac{\partial u_i}{\partial x_j} + \frac{\dd u_j}{\dd x_i} \right),\\
S & = & \frac{1}{2} \sigma_{kk}, \\ 
E & = & \frac{1}{2} \epsilon_{kk},\\
s_{ij} & = & \sigma_{ij} - \delta_{ij} S, \\
e_{ij} & = & \epsilon_{ij} - \delta_{ij} E,\\
G_{\tau} \dot{s}_{ij} + s_{ij} & = & G_2 G_\tau \dot{e}_{ij} + G_1 e_{ij}, \\
K_{\tau} \dot{S} + S & = & K_2 K_\tau \left( \dot{E}- \BB \dot{\beta} \right) + K_1 (E- \BB \beta),\\
\label{msend}  \frac{\partial}{\partial t} \left( \frac{1}{\Delta} \frac{\partial u_i}{\partial X_i}+ \frac{\partial u_i}{\partial x_i} \right) & = &  \frac{1}{\Delta^2} \frac{\partial^2 p}{\partial X_i \partial X_i}+ \frac{2}{\Delta} \frac{\partial^2 p}{\partial X_i \partial x_i} + \frac{\partial^2 p}{\partial x_i \partial x_i} .
\ee
The boundary conditions on the inclusion (\ref{cheesy})--(\ref{feta}) become
\be
\label{msBC} u_i = \Delta X_i {\mathcal G} g + U_i +\Delta \Theta \varepsilon_{ij3}X_j \quad \text{and} \quad {\left( \frac{1}{\Delta} \frac{\dd p}{\dd X_i} + \frac{\dd p}{\dd x_i}\right) n_i = 0} \on \ginc,
\ee
where, $\ginc$ is the circle centered at the origin of the microscopic coordinate $X_i$ with dimensionless radius $R<1/2$, {cf.~Fig.~\ref{fig:figure1}}. {The nondimensional form of \ref{constraintnd} is 
\be \label{fish}
\int_{ \Gamma_{inc}} \left( \sigma_{ij} {- \PP \delta_{ij} p} \right) {n}_j \, d\Gamma =0 \quad \mbox{and} \quad \int_{ \Gamma_{inc}} \varepsilon_{ik3} x_k \sigma_{ij} n_j \, d\Gamma =0.
\ee
Some care is needed again to put these integral constraints in multiple scales form \cite{Cha15}. We defer the discussion until later.
}

{We now expand each of the dependent variables in powers of $\Delta$ as} 
\be \label{laughing}
y = \yo(\bx,\bX) + \Delta \yone(\bx,\bX) + O(\Delta^2).
\ee


\subsection{The leading order problem}
{At leading order in (\ref{ms1b}), (\ref{msBC}a) and (\ref{new}a) we find}
\be \label{lo}
\frac{\partial \uo_i}{\partial X_j} + \frac{\dd \uo_j}{\dd X_i} = 0, \quad \frac{\partial \Uo_i}{\partial X_j}=0,  \quad  \mbox{with} \quad \uo_i = \Uo_i \on \Gamma_{inc}.
\ee
{The solution to (\ref{lo}) is a rigid body motion; periodicity on the microscale requires that this motion is a pure translation. Thus $\uo_i$ and $\Uo_i$ are independent of $\bX$ and 
\be \label{uop}
\uo_i(\bx,t) = \Uo_i(\bx,t). 
\ee}
{At leading order  in (\ref{msend}) and (\ref{msBC}b) we find}
\be
{\frac{\partial^2 \po}{\partial X_i \partial X_i }=0} \quad \text{with} \quad \frac{\dd \po}{\dd X_i} n_i = 0 \on \Gamma_{inc}
\ee
{which together with {periodicity in $\bX$} implies} that 
\be \label{lop}
{\po = \po(\bx,t).}
\ee
Finally, at leading order in (\ref{ms1}), using (\ref{lop}) we find
\be \label{pushed}
\frac {\dd \sigmao_{ij}}{\partial X_j} =0;
\ee
although this is not sufficient to determine the leading-order stress we will use it in the next section as part of the problem to be solved at first order.


\subsection{The first order problem}\label{FOP}
{Equating coefficients of powers of $\Delta$ at the next order in (\ref{new})--(\ref{msBC}) we find that}
\be
\label{la} \epsilono_{ij} & = & \frac{1}{2} \left( \frac{\partial \uone_i}{\partial X_j} + \frac{\dd \uone_j}{\dd X_i} \right) + {\frac{1}{2} \left( \frac{\partial \uo_i}{\partial x_j} + \frac{\dd \uo_j}{\dd x_i} \right)},\\
\So & = & \frac{1}{2} \sigmao_{kk}, \\
\Eo & = & \frac{1}{2} \epsilono_{kk},\\
\so_{ij} & = & \sigmao_{ij} - \delta_{ij} \So, \\
\eo_{ij} & = & \epsilono_{ij} - \delta_{ij} \Eo, \\
\label{foend} G_{\tau} \dot{s}\oo_{ij} + \so_{ij} & = & G_2 G_\tau \dot{e}\oo_{ij} + G_1 \eo_{ij}, \\ \label{sniff} K_{\tau} \dot{S}\oo + \So & = & K_2 K_\tau \left( \dot{E}\oo - \BB {\dot{\beta}} \right) + K_1 (\Eo - \BB {\beta}),\\
\label{shuffle}
{\frac{\dd^2 \pone}{\dd X_i \dd X_i}} & {=} & {0},\\
\label{bridge} \frac{\partial \Uone_i}{\partial X_j} + \frac{\partial \uo_i}{\partial x_j}&=&0, \qquad \frac{\dd \Tho}{\dd X_j}=0, 
\ee
{with}
\be
\label{foBC} \uone_i = X_i {\mathcal G} {g} + \Uone_i +\Tho \varepsilon_{ij3} X_j\quad \mbox{and} \quad { \left( \frac{\dd \pone}{\dd X_i} + {\frac{\dd \po}{\dd x_i}} \right) {n}_i = 0,} \on \Gamma_{inc},
\ee
where terms involving the leading-order pressure and inclusion displacements vanish because they are independent of the microscopic scale; see (\ref{uop}) and (\ref{lop}). {In addition, the leading-order behaviour of the integral constraints (\ref{fish}) is simply}
\be
\label{flap} \int_{\Gamma_{inc}} \varepsilon_{ik3} \sigmao_{ij} X_k n_j \, d\Gamma = 0, \qquad
\int_{\Gamma_{inc}} \sigmao_{ij} \, d\Gamma = 0,
\ee
{and all variables are periodic in $\bX$ with period one.}
On supplementing (\ref{la})--(\ref{flap}) with the as yet unused equations (\ref{pushed}) we have a closed problem for seventeen unknowns $\uone_i$, $\sigmao_{ij}$, $\epsilono_{ij}$, $\So$, $\Eo$, $\so_{ij}$, $\eo_{ij}$, $\pone$ and $\Uone_i$. The problem is excited by four different forms of forcing, namely:
\begin{enumerate}[(i)]
\item leading order macroscopic deformation gradient, $\partial \uone_i/\partial x_j$, which appears in both (\ref{la}) and (\ref{bridge});
\item the eigenstrain induced by binder swelling, $ \beta$, which appears in (\ref{sniff}), 
\item the deformation induced by the volumetric changes of the electrode particles, $g(t)$, which appears in (\ref{foBC}a); {and}
\item the leading order macroscopic pressure gradient, $\partial \po/\partial x_i $, which appears in (\ref{foBC}b).
\end{enumerate}
Crucially, the system (\ref{la})--(\ref{flap}) is linear and therefore
we can solve by finding four ``cell
functions'', each of which represents the response of the system to a
unit forcing of one of the four modes described above.  It is worth
noting that {since $p^{(1)}$ only appears in (\ref{shuffle}) and (\ref{foBC}b), and these equations decouple from the rest}, 
the fluid and solid responses are decoupled. Thus forcings (i)--(iii) do not excite the fluid, whilst forcing (iv) generates no solid deformation or stress. {Denoting the cell functions corresponding to the each of the modes (i)--(iv) by $\sigma^{kl}_{ij}$, $\sigma^{\beta}_{ij}$, $\sigma^{g}_{ij}$ and $p^q$ 
respectively, we have, in general, }
\begin{multline}
\label{totals}
\sigmao_{ij} =  \int_0^t \sigma^{kl}_{ij}(X_i,t-t') \frac{\partial \uo_k}{\partial x_l}(t') \, dt' \\ 
+  \int_0^t \sigma^\beta_{ij}(X_i,t-t') {\mathcal B}\left( K_2 K_\tau \dot{\beta}(t') + K_1 \beta(t') \right)\, dt'\\
+  \int_0^t\sigma^{g}_{ij}(X_i,t-t') {\mathcal G} g(t') \,dt',
\end{multline} 
and
\be \label{press}
\pone = p^{q}(X_i) \frac{\dd \po}{\dd x_q}.
\ee 
Finding the four cell functions ($\sigma^{kl}_{ij}$, $\sigma^{\beta}_{ij}$, $\sigma^{g}_{ij}$ and $p^q$) is a task that must be tackled numerically and rather than interrupting the homogenisation process to describe how this has been done we {defer} these details to appendices \ref{macstr}-\ref{mpg}.

\subsection{Effective momentum balance} Proceeding one additional order in the expansion of (\ref{ms1}) gives
\be
\label{fo1} \frac{\dd}{\dd X_i} \left( \sigmaone_{ij} - \PP \delta_{ij} \pone \right) + \frac{\dd}{\dd x_i} \left( \sigmao_{ij} - \PP \delta_{ij} \po \right)  = 0. 
\ee
On integrating (\ref{fo1}) over the {unit cell} and applying the divergence theorem we arrive at
\begin{multline}
- \int_{\Gamma_\text{inc}} (\sigmaone_{ij} - \PP \delta_{ij} \pone) n_j \, d\Gamma
+\int_{\Gamma_\text{box}} (\sigmaone_{ij} - \PP \delta_{ij} \pone) n_j \, d\Gamma \\
 + \frac{\dd}{\dd x_j} \left( \int_{\Omega} \sigmao_{ij} - \PP \delta_{ij} \po \, d\Omega \right) = 0,
\end{multline}
where the  order of integration and differentiation in the {final} term {has been exchanged, which is allowed because of} the multiple scales assumption that $x_i$ and $X_i$ are independent.
Note that the minus sign in the first term arises because of our definition of the normal vector, which points outwards from the inclusion and into the matrix; see~Fig.~\ref{fig:figure1}. {Periodicity implies that the integral over  $\Gamma_{\text{box}}$ vanishes, leaving}
\be \label{millen}
-\int_{\ginc} (\sigmaone_{ij} -\PP \delta_{ij} \pone) n_j d\Gamma + \frac{\dd}{\dd x_j} \left( \int_{\Omega} \sigmao_{ij} -\PP \delta_{ij} \po d\Omega \right) = 0.
\ee
It is tempting to conclude, on the basis of (\ref{fish}), that the first term is zero. However, it has been shown in \cite{Cha15} that this is not the case. Instead we have that 
\be \label{jon}
-\int_{\ginc} \left( \sigmaone_{ij} - \PP \delta_{ij} \pone \right)  n_j d\Gamma = \int_{\ginc} X_j \frac{\dd}{\dd x_j} \left( \sigmao_{ik} - \PP \delta_{ik} \po \right) n_k d\Gamma;
\ee
{the right-hand side arises from the small variation in $x_i$ around $\ginc$, and is the correct way to put the integral constraint into multiple scales form}. Substituting (\ref{jon}) into (\ref{millen}) we arrive at 
\be \label{philip}
\frac{\dd}{\dd x_j} \left(\int_{\Omega} \sigmao_{ij} - \PP \delta_{ij} \po d\Omega + \int_{\ginc} X_j (\sigmao_{ik} - \PP \delta_{ik} \po) n_k d\Gamma \right) =0.
\ee
{Since $\po$ is independent of $\bX$,
\begin{eqnarray*}
\int_{\Omega} \PP \delta_{ij}\po  d\Omega & = & \PP \po |\Omega| \delta_{ij},\\
\int_{\ginc} X_j  \PP \delta_{ik} \po n_k d\Gamma &=&\PP \po \int_{\ginc} X_j   \delta_{ik}  n_k d\Gamma =  
  \PP \po \int_{\Omega'} \frac{\dd}{\dd X_k}(X_j   \delta_{ik})\, d \Omega =
  \PP \po |\Omega'| \delta_{ij}.
\end{eqnarray*}
Since $\sigmao_{ij}$ depends on $\bX$ we cannot use the same trick on the other surface integral, but by defining the effective stress (and averaging operator ${\cal A}$) as 
\begin{equation}
\sigmaeff_{ij} = {\mathcal A} \left( \sigmao_{ij} \right) = \frac{1}{|\Omega+\Omega'|}\int_{\Omega} \sigmao_{ij} \, d\Omega + \frac{1}{|\Omega+\Omega'|}\int_{\ginc} X_j \sigmao_{ik}  n_k \, d\Gamma ,\label{averagingOp}
\end{equation}
(\ref{philip}) reads
\be \label{this1}
{\frac{\dd}{\dd x_j} \left( \sigmaeff_{ij} - \PP \delta_{ij} \peff  \right) = 0,}
\ee
where  $\peff = \po$.}
This is the momentum balance equation for the effective model.
{Note that the torque condition (\ref{flap}) implies that
\be \label{braddd}
\int_{\ginc}  X_1 \sigma_{2k}^{(0)} n_k d\Gamma= \int_{\ginc}  X_2 \sigma_{1k}^{(0)} n_k d\Gamma,
\ee
so that $\sigmaeff$ is symmetric, as expected.
}

\subsection{Effective Darcy flow in a deformable medium}
{Proceeding one order further in the expansion of (\ref{msend}) and (\ref{msBC}b) gives}
\be
\label{balls}\frac{\dd}{\dd X_i} \left( \frac{\dd \ptwo}{\dd X_i} + \frac{\dd \pone}{\dd x_i} - \frac{\dd \uone_i}{\dd t} \right) + \frac{\dd}{\dd x_i} \left( \frac{\dd \pone}{\dd X_i} + \frac{\dd \po}{\dd x_i} - \frac{\dd \uo_i}{\dd t} \right)  &=& 0,\\
\label{tackle}\left( \frac{\dd \ptwo}{\dd X_i} + {\frac{\dd \pone}{\dd x_i}} \right) {n}_i &=& 0, \on \Gamma_{inc},
\ee
{Integrating (\ref{balls})  over $\Omega$ and applying the divergence theorem and (\ref{tackle})  leads to}
\be \label{fuckpiss}
\int_{\ginc} \frac{\dd \uone_i}{\dd t} n_i d\Gamma + \frac{\dd}{\dd x_i} \left( \int_{\Omega} \frac{\dd \pone}{\dd X_i} + \frac{\dd \po}{\dd x_i} - \frac{\dd \uo_i}{\dd t} d\Omega \right) =0,
\ee
{where, as before, contributions from $\Gamma_\text{box}$ vanish due to periodicity.
Using (\ref{foBC}a) and (\ref{bridge}) we find
\be 
\nonumber\int_{\ginc} \frac{\dd \uone_i}{\dd t} n_i d\Gamma &=& \int_{\ginc} \left( X_i {\mathcal G} \dot{g} + \frac{\dd \Uone_i}{\dd t} + \frac{\dd \Tho}{\dd t} \epsilon_{ij3} X_j \right) n_i d\Gamma,\\
\nonumber &=& \int_{\ginc} \left( X_i {\mathcal G} \dot{g} - X_j \frac{\dd^2 \uo_i}{\dd x_j\dd t} + \frac{\dd \Tho}{\dd t} \epsilon_{ij3} X_j \right) n_i d\Gamma,\\
&=&|\Omega'|\left(2{\mathcal G} \dot{g} - \frac{\dd^2 \uo_i}{\dd x_i\dd t}\right) .
\label{sick}
\ee
(The 2 here arises because we are in two spatial dimensions---in $d$ spatial dimensions the 2 becomes a $d$).
Now using  (\ref{press}) in (\ref{fuckpiss}) gives
\be \label{fuckpiss2}
|\Omega'|\left(2{\mathcal G} \dot{g} - \frac{\dd^2 \uo_i}{\dd x_i\dd t}\right)
   + \frac{\dd}{\dd x_i} \left(  \frac{\dd \po}{\dd x_q} \int_{\Omega} \frac{\dd p^q}{\dd X_i} d\Omega + |\Omega| \frac{\dd \po}{\dd x_i} - |\Omega| \frac{\dd \uo_i}{\dd t}  \right) =0.
\ee}
If we define {the effective permeability as}
\be\label{kappaeff}
{\kappa}^{\text{eff}}_{iq} =  \frac{1}{|\Omega+\Omega'|} {\displaystyle \int_{\Omega} \delta_{iq} + \frac{\dd p^q}{\dd X_i} d\Omega},
\ee
{then,} equation (\ref{fuckpiss2}) may be written as
{\be \label{dumdung}
2 {(1-\phi)} {\mathcal G} \dot{g} + \frac{\dd}{\dd x_i} \left( {\kappa}^{\text{eff}}_{iq} \frac{\dd \peff}{\dd x_q}  - \frac{\dd \ueff_i}{\dd t} \right) =0.
\ee}
{where $\ueff_i=\uo_i$} and 
\be
\phi = \frac{|\Omega|}{|\Omega+\Omega'|},
\ee
is the volume fraction of porous binder and electrolyte (the light blue region in Figs~\ref{fig:figure1} and \ref{RVEs}).
{The additional source term in the effective Darcy equation (\ref{dumdung}) accounts for the fluid that is displaced by the growth/shrinkage of the inclusion.}


\subsection{Effective constitutive relation}
{To close the homogenised model we need to determine the effective constitutive relation between $\sigmaeff$ and ${\bf u}^{\text{eff}}$. Applying the averaging operator ${\cal A}$ in (\ref{averagingOp}) to (\ref{totals}), recalling that ${\bf u}^{\text{eff}}$ is independent of $\bX$, gives this constitutive relation as}
\begin{multline} \label{break} 
\sigmaeff_{ij} =  \int_0^t \sigma^{kl,\text{eff}}_{ij}(t-t') \frac{\dd \ueff_k}{\dd x_l} (t') \, dt'  
+  \int_0^t \sigma^{\beta,\text{eff}}_{ij}(t-t') {\mathcal B}\left( K_2 K_\tau \dot{\beta}(t') + K_1 \beta(t') \right)\, dt'\\
+  \int_0^t\sigma^{g,\text{eff}}_{ij}(t-t') {\mathcal G} g(t') \,dt'.
\end{multline}  
where $\sigma^{kl,\text{eff}}_{ij} = {\mathcal A} ( \sigma^{kl}_{ij} )$, $\sigma^{\beta,\text{eff}}_{ij} = {\mathcal A} ( \sigma^\beta_{ij} )$, $\sigma^{g,\text{eff}}_{ij} = {\mathcal A} ( \sigma^g_{ij} )$.
 In general, the effective material is anisotropic, however, as we shall see shortly, the constitutive {relations simplify significantly when the RVE has certain symmetries.}

\paragraph{Symmetric RVEs}
{The microstructure we focus on has four-fold symmetry (see Fig.~\ref{fig:figure1}). This, along with the symmetries of the stress and strain tensor, implies
\be\label{biscuit}
\sigma^{11,\text{eff}}_{11} = \sigma^{22,\text{eff}}_{22}, \qquad \sigma^{11,\text{eff}}_{22} = \sigma^{22,\text{eff}}_{11}, \qquad \sigma^{12,\text{eff}}_{12} = \sigma^{21,\text{eff}}_{21} = \sigma^{12,\text{eff}}_{21} = \sigma^{21,\text{eff}}_{12}, \label{skin} \\
\sigma^{12,\text{eff}}_{11} = \sigma^{21,\text{eff}}_{22}=  \sigma^{12,\text{eff}}_{22} = \sigma^{21,\text{eff}}_{11}=   \sigma^{11,\text{eff}}_{12} = \sigma^{22,\text{eff}}_{21} =  \sigma^{11,\text{eff}}_{21} = \sigma^{22,\text{eff}}_{12}=0.\label{cheese}
\ee
Thus the effective material is orthotropic, with the three independent components of the response arising from forcing mode (i)}.
{The responses to forcing modes (ii) and (iii) are purely volumetric, so that these cell functions can be written} 
\be \label{gilbert}
\sigma^{\beta,\text{eff}}_{ij} = \delta_{ij} S^{\beta,\text{eff}} \qquad \text{and} \qquad 
\sigma^{g,\text{eff}}_{ij} = \delta_{ij} S^{g,\text{eff}},
\ee
with each having one independent component.
This motivates separating the effective constitutive equation into deviatoric and volumetric parts, and to this end we introduce 
\be \label{easter}
\Seff = \frac{1}{2} \sigmaeff_{kk}, \quad \seff_{ij} = \sigmaeff_{ij} - \delta_{ij} \Seff, \quad S^{kl,\text{eff}} = \frac{1}{2} \sigma^{kl,\text{eff}}_{mm}, \quad s^{kl,\text{eff}}_{ij} = \sigma^{kl,\text{eff}}_{ij} - \delta_{ij} S^{kl,\text{eff}},\\
\Eeff = \frac{1}{2} \epsiloneff_{kk}, \quad \eeff_{ij} = \epsiloneff_{ij} - \delta_{ij} \Eeff, \quad E^{kl,\text{eff}} = \frac{1}{2} \epsilon^{kl,\text{eff}}_{mm}, \quad e^{kl,\text{eff}}_{ij} = \epsilon^{kl,\text{eff}}_{ij} - \delta_{ij} E^{kl,\text{eff}}, \label{bunny}
\ee
where 
\be \label{prof}
\epsiloneff_{ij} = \frac{1}{2} \left( \frac{\partial \ueff_i}{\partial x_j} + \frac{\partial \ueff_j}{\partial x_i} \right).
\ee
Equations (\ref{skin})--(\ref{prof}) allow us to rewrite (\ref{break}) as
\be
{\seff_{11} = \int_0^t 2 s^{11,\text{eff}}_{11} (t-t') \eeff_{11}(t') ~dt',}\\
{\seff_{12} = \int_0^t 2 s^{12,\text{eff}}_{12} (t-t') \eeff_{12}(t') ~dt',}\\
\begin{aligned}
\Seff = \int_0^t & S^{\beta,\text{eff}}(t-t') {\mathcal B} \left( K_2 K_\tau \dot{\beta}(t')+K_1 \beta(t') \right) + S^{g,\text{eff}}(t-t') {\mathcal G} g(t') \\ + & 2 S^{11,\text{eff}}(t-t') \Eeff(t') ~dt'. 
\end{aligned} \label{grubs}
\ee
{Thus the composite behaves as an orthotropic linear viscoelastic
  material, but with a more complex rheology than the constituent
  binder}. Fig.~\ref{Cells} summarises the constitutive behaviour of
the effective material where we show the Laplace transform (indicated
with a hat) of the functions charactersising the effective response
($s^{11,\text{eff}}_{11}$, $s^{12,\text{eff}}_{12}$,
$S^{11,\text{eff}}$, $S^{\beta,\text{eff}}$ and $S^{g,\text{eff}}$) as
a function of {the transform variable $w$. These functions were
mapped out across the  complex
$w$-plane, but the interesting features lie along the purely real
axis}. The presence of a singularity corresponds to a relaxation
timescale in the constitutive behaviour in the time domain. Thus, we
can see that the effective material inherits the bulk and shear
relaxation timescales ($K_{\tau}$ and $G_{\tau}$) from the constituent
binder, but both of these timescales appear in both the effective bulk
and shear responses. Additionally the effective material gains four
additional relaxation scales, labelled {$G_\tau^1$--$G_\tau^4$}.  


{The observation that the homogenised response is dominated by a
handful of discrete timescales suggests  an {heuristic}
approach to characterising the effective material in which a
generalised Maxwell model is fitted to the functions shown in
Fig.~\ref{Cells}. This is likely to yield accurate constitutive
behaviour and be not overly onerous to implement in
computations. 
We have carried out some simulations on RVEs with less symmetry than
that shown in Fig.~\ref{fig:figure1} and it appears that this feature,
of discrete relaxation timescales being introduced by homogenisation,
remains robust.}

\subsection{Summary of the effective equations}
The statement of the effective model for the microscale geometry shown
in Fig.~\ref{fig:figure1} can now be formulated. Equation
(\ref{this1}) provides the effective momentum balance, (\ref{break})
is the effective constitutive relation, the effective equation for
Darcy flow in a deformable medium is given by (\ref{dumdung}),  and (\ref{prof}) is the definition of the effective strain. In summary the effective system is
\be
\label{1effa} \frac{\partial }{\partial x_j} \left( \sigmaeff_{ij} {- {\mathcal P} \delta_{ij} \peff} \right) & = & 0, \\
\label{1effb}\epsiloneff_{ij} & = & \frac{1}{2} \left( \frac{\partial \ueff_i}{\partial x_j} + \frac{\dd \ueff_j}{\dd x_i} \right),\\
\label{2eff} \Seff & = & \frac{1}{2} \sigmaeff_{kk}, \\
\Eeff & = & \frac{1}{2} \epsiloneff_{kk},\\
\seff_{ij} & = & \sigmaeff_{ij} - \delta_{ij} \Seff, \\
\label{3eff}\eeff_{ij} & = & \epsiloneff_{ij} - \delta_{ij} \Eeff,\\
\label{4eff} \seff_{11} & = & \int_0^t 2 s^{11,\text{eff}}_{11} (t-t') \eeff_{11}(t') ~dt',\\
\seff_{12} & = & \int_0^t 2 s^{12,\text{eff}}_{12} (t-t') \eeff_{12}(t') ~dt',\\
\label{twiglet} 
\Seff & = & \int_0^t S^{\beta,\text{eff}}(t-t') {\mathcal B} \left(
  K_2 K_\tau \dot{\beta}(t')+K_1 \beta(t') \right) \nonumber \\  & &
\mbox{ }+ S^{g,\text{eff}}(t-t') {\mathcal G} g(t')  + 2 S^{11,\text{eff}}(t-t') \Eeff(t') ~dt',\\
\label{effend} 2 {(1-\phi)} {\mathcal G} \dot{g} & = & -\frac{\dd}{\dd x_i} \left( \kappa^{\text{eff}}_{ij} \frac{\dd \peff}{\dd x_j}  - {\phi}\frac{\dd \ueff_i}{\dd t} \right).
\ee
These equations are to be solved subject to suitable boundary and initial conditions (examples given in the subsequent section). Contrasting the effective equations, (\ref{1effa})--(\ref{effend}), with the original model, (\ref{gov1nd})--(\ref{constraintnd}), we see that the two are broadly similar, but with some important differences{: (i)} the homogenised constitutive equations capture the effects of particle and binder swelling as sources of volumetric stress; (ii) a (finite) number of additional viscoelastic relaxation timescales are introduced by the homogenisation, as discussed in more detail below (\ref{grubs}), and; (iii) the effective Darcy law contains a source/sink of fluid which accounts for the fluid motion driven by the growth/shrinkage of particles that removes/adds pore space in within the binder.

\begin{figure}
\centering
\includegraphics[scale=0.45]{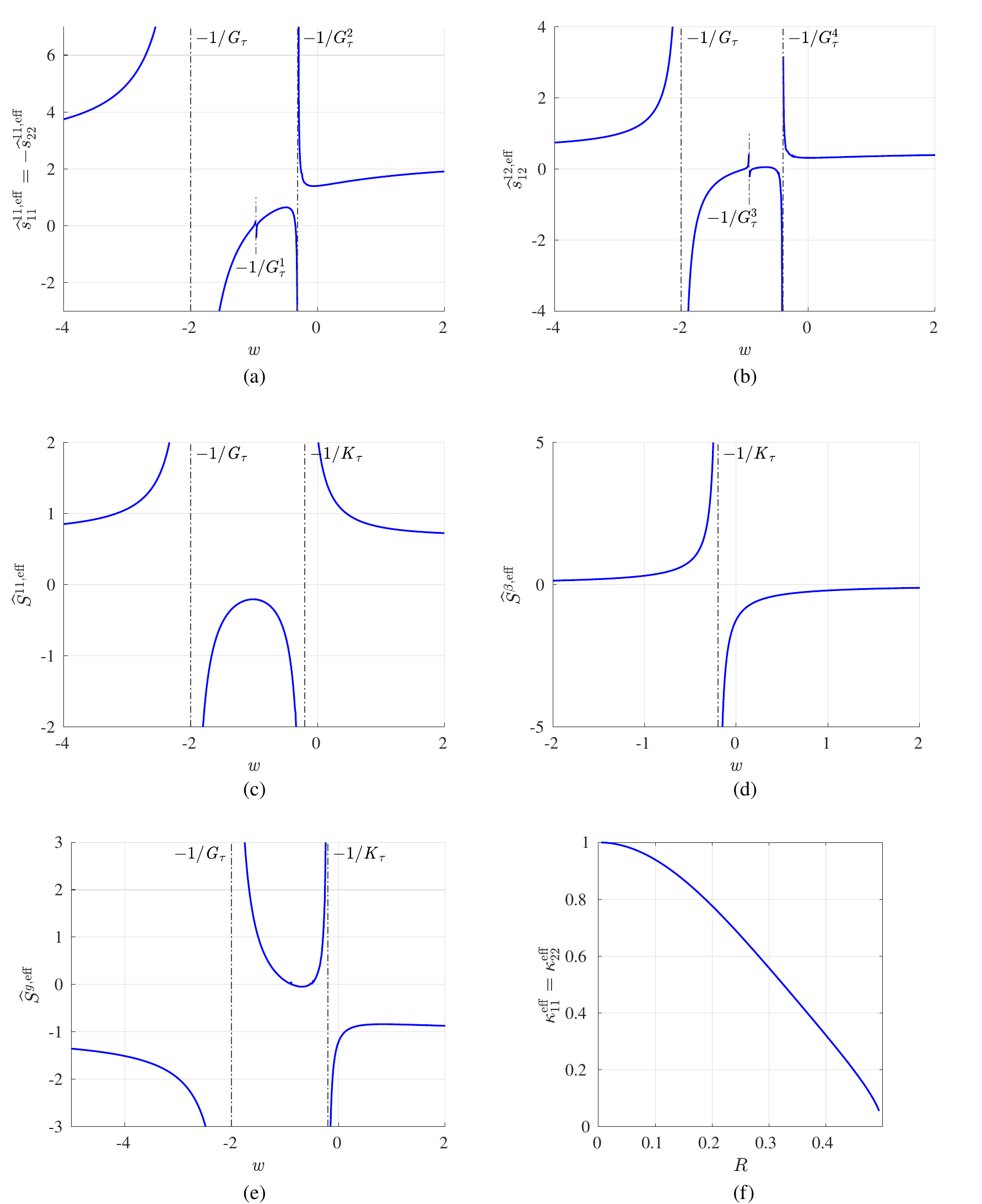}
\caption{Summary of the cell functions which appear in the effective model. Panels (a)--(c) characterise the response to macroscopic deformation gradients, whilst (d) and (e) characterise the response to binder swelling and active particle swelling respectively, and (f) characterises the response of the effective permeability. }
\label{Cells}
\end{figure}

\paragraph{Comparison to  prior approaches}
{In \cite{Fos17} a similar problem to the present one was 
  considered. There, solutions were sought via ad-hoc homogenisation
  rather than via the systematic multiple scale approach used
  here. The microscopic governing equations were solved on an RVE
  choosing boundary conditions consistent with a particularly simple
  one-dimensional macroscopic solution (relevant because
  battery electrodes typically have extremely high aspect ratios; see
  \S\ref{results}). Comparing our
  equations (\ref{1effa})--(\ref{effend}) to (3.1)--(3.18) in
  \cite{Fos17} we find that the ad-hoc approach correctly
  {captures} the majority of the features of the one-dimensional
  macroscopic solution, but 
  it fails to account for the complex rehological features of the
  effective material (the effective material in \cite{Fos17} has
  identical relaxation timescales as the constituent binder). Given
  the lack of a consensus on the correct viscoelastic characterisation
  for the relevant polymers one could argue that this is not an
  important deficiency because, in practice, one will likely require
  experimental characterisation and fitting to parameterise a
  model. However, in addition to being limited to one-dimensional
  macroscopic deformations, the ad-hoc approach also does not 
  account for the flow of the electrolyte. }

\section{\label{results}Representative simulations and validation}
We will now solve both the effective and full models in three
different scenarios, each of which is relevant to real
batteries. Lithium-ion battery electrodes are slender with thicknesses
in the direction normal to the current collectors usually being
$L_2 \sim 100\:\mu$m whereas their lateral dimensions are often $L_1\sim 10$ cm;
see Fig.~\ref{fig:figure1}. The slender and planar geometry can be
exploited by scaling $x_1$ with $L_1/L_2$ where $L_2/L_1 \ll 1$ and
seeking an asymptotic solution in the limit that $L_2/L_1\to 0$.  In
the interests of brevity, we 
{do not make the expansions explicit}, and instead we point out
that in the remainder of this section all variables should be
understood to be leading order approximations only. The extreme aspect
ratio of the electrode means that throughout the bulk region (away
from the lateral extremities) the leading order deformation is
particularly simple and has the form 
\be
\label{defform} \ueff_1 = 0, \qquad \ueff_2 = \ueff_2(x_2,t).
\ee 
From which it follows, via (\ref{1effb}b)--(\ref{3eff}b), that
\be
\label{defeps} \epsiloneff_{11} = \epsiloneff_{12} = \eeff_{12} = 0, \quad \epsiloneff_{22} = \frac{\partial \ueff_{2}}{\partial x_2}, \quad \eeff_{11} = -\frac{1}{2} \frac{\partial \ueff_{2}}{\partial x_2}, \quad \eeff_{22} = \frac{1}{2} \frac{\partial \ueff_{2}}{\partial x_2}, \quad \Eeff = \frac{1}{2} \frac{\partial \ueff_{2}}{\partial x_2}.
\ee
The leading order terms in (\ref{1effa}a) and (\ref{effend}) are
\be \label{snorton}
\frac{\dd}{\dd x_2} \left( \sigmaeff_{12} \right) = 0, \:\:
\frac{\dd}{\dd x_2} \left( \sigmaeff_{22} - \PP \peff \right) = 0,
\quad {2{(1-\phi)}} {\mathcal G} \dot{g} + \frac{\dd}{\dd x_2} \left(
  \kappa^{\text{eff}}_{22} \frac{\dd \peff}{\dd x_2} - {\phi}\frac{\dd
    \ueff_2}{\dd t} \right) = 0, 
\ee
{whilst the remaining (constitutive) equations, (\ref{4eff})--(\ref{twiglet}), retain all terms at leading order. {We emphasize that the reduced effective equations \eqref{defeps}--\eqref{snorton} and their solutions are independent of the coordinate $x_1$.}
We will now proceed to study the solution in three scenarios, namely; calendering, cycling and impact. Details on the implementation of the simulations are presented in appendix \ref{implt}.}

\begin{table}{
\centering
        \begin{tabular}{ccccccc}
        \toprule[1.2pt]
        $G_\tau$ & $G_1$ & $G_2$ & $K_\tau$ & $K_1$ & $K_2$ & $\alpha$ \\
        \midrule\midrule
		$1/2$ & $2$ & $4$ & $5$ & $3$ & $1/3$ & 0.25  \\
		\bottomrule[1.2pt]
        \end{tabular}
\caption{Dimensionless parameters for simulations throughout \S\ref{results}.}}\label{Tab1}
\end{table}

\subsection{Calendering}
Calendering is a manufacturing step in which dry electrodes are briefly compressed using a large-radius steel roller~\cite{Gim19,Gim20,Meyer2017}. Since the electrode is dry, we set ${\mathcal P}=p=0$. Appropriate boundary conditions are
\be
\ueff_1 = \ueff_2 = 0 \quad \text{on} \quad x_2 = 0,\\
\ueff_1 = 0, \quad \ueff_2 = u_2^{\text{app}}(t) \quad \text{on} \quad x_2 = 1,\\*[2mm]
\label{bigbarry} \text{where} \quad {u_2^{\text{app}}(t) = \left\{\begin{array}{cl}
	-2.0415 t & 0 < t \leq 1/3\\
	-0.6805 & 1/3 < t \leq 2/3\\
	0.9585t-1.3195 & 2/3 < t \leq 1
\end{array}\right.}
\ee
where ${x_2}=0$ corresponds to the rigid current collector and ${x_2}=1$ the upper surface in contact with the roller. The displacement on the top surface mimicks (loosely) the deformation induced on the electrode as it passes under the roller. The size and timescale of the applied deformation has been chosen to approximately match the experiments conducted in \cite{Gim19,Gim20}. We shall assume stress- and strain-free initial state such that
\be \label{usually}
E|_{t=0} = e_{ij}|_{t=0} = S|_{t=0} = s_{ij}|_{t=0} = 0.
\ee
The mechanical and geometrical parameters are summarised in Table~\ref{Tab1} and since there is no electrolyte (the binder does not swell and the particles are electrochemically inactive) we take $\mathcal{G}=\mathcal{B}=0$.

The effective vertical deformation is given by 
\be 
\ueff_2 = {x_2} u_2^{\text{app}}(t),
\ee
which, on back-substitution into \eqref{defeps}, determines all non-zero effective strain components. In turn these can be substituted into \eqref{4eff} and \eqref{twiglet} to determine the non-zero components of the effective stress
\be
\label{prelap4} s^{\text{eff}}_{11} = -\int_0^t s^{11,\text{eff}}_{11}(t-t') \ueff_2(t') \, dt', \quad \Seff = \int_0^t S^{11,\text{eff}}(t-t')\ueff_2(t') ~dt'.
\ee
Subtracting the former equation in (\ref{prelap4}) from the latter, and taking a Laplace transform yields
\be
\hat{\sigma}_{22}^{\text{eff}} = \left(\hat{s}^{11,\text{eff}}_{11} + \hat{S}^{11,\text{eff}}\right)\hat{u}_2^{\text{app}}.\label{result1}
\ee

In Fig.~\ref{calendering1} we compare (\ref{result1}) to the corresponding solution of the full model for a variety of values of $\Delta$. {In the numerical simulations of the full model we set ${\mathcal P}=p=\mathcal{G}=\mathcal{B}=0$, and the boundary conditions are
\be
u_i = 0 &\text{on} & x_2 = 0,\\
u_1 = 0, \quad \text{and} \quad u_2 = u_2^{\text{app}}(t) &\text{on}& x_2 = 1,\\
u_i, \quad \text{and} \quad \sigma_{ij}n_j \quad \text{are periodic in }
  & x_1 & \text{ with period $\Delta$.}   
\ee}
As expected, we observe high compressive stresses between the
particles due to the loading from the roller. As $\Delta$ decreases, the
solution to full model converges to the homogenised counterpart, as
corroborated by Fig.~\ref{calendering2}. Other simulations (not shown)
with different parameter values show similarly good convergence.

{A key point about our homogenisation procedure is that the leading order micro-resolved quantities are not lost. Rather they can be recovered having solved the homogenised model. Together, the Laplace transform of (\ref{totals}) and (\ref{result1}) allow us to write
\be
\hat{\sigma}_{22}^{(0)} = \left( \hat{s}_{22}^{22} + \hat{S}^{22}\right)u_2^{\rm app} \label{laughinggas}
\ee 
thereby recovering the leading order micro-resolved stress from the effective solution. A comparison between the leading order stress according to (\ref{laughinggas}) and the stress computed by direct numerical simulation of the full model is shown in Fig.~\ref{laughinggasfig} where excellent agreement is observed.}

{To demonstrate the homogenised model's prediction in the time-domain we invert the Laplace transform in (\ref{laughinggas}), yielding the leading-order 22-component of the stress. This quantity evaluated at specific spatial points, as a function of time, is shown in Fig.~\ref{mrweewee}. The behaviour appears almost elastic during the application ($0<t<1/3$) and during the release ($0<t<1/3$) of the load. When the applied deformation is constant ($1/3<t<2/3$) the expected viscoelastic relaxation is observed.}
	
	


\begin{figure}
\centering
\includegraphics[scale=0.9]{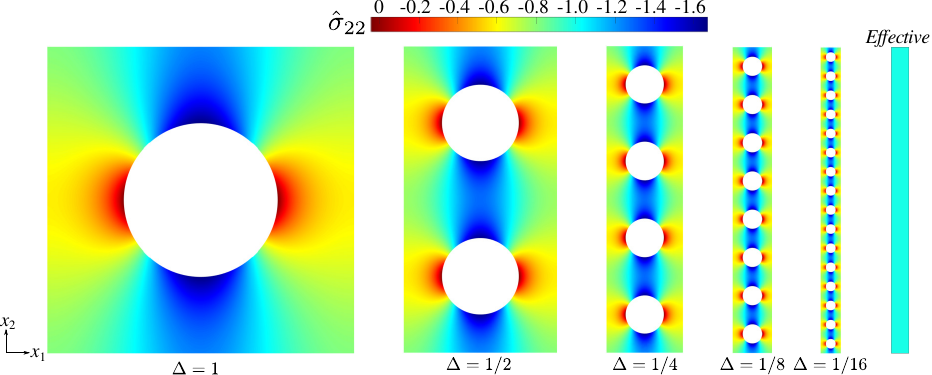}
\caption{{Predictions of $\hat{\sigma}_{22}$ at
  $w = 0.505$ (where $w$ is the transform variable) for the
  calendering problem by direct numerical simulations vs. the counterpart effective quantity according to (\ref{result1}).} } 
\label{calendering1}
\end{figure}

\begin{figure}
\centering
\includegraphics[scale=0.9]{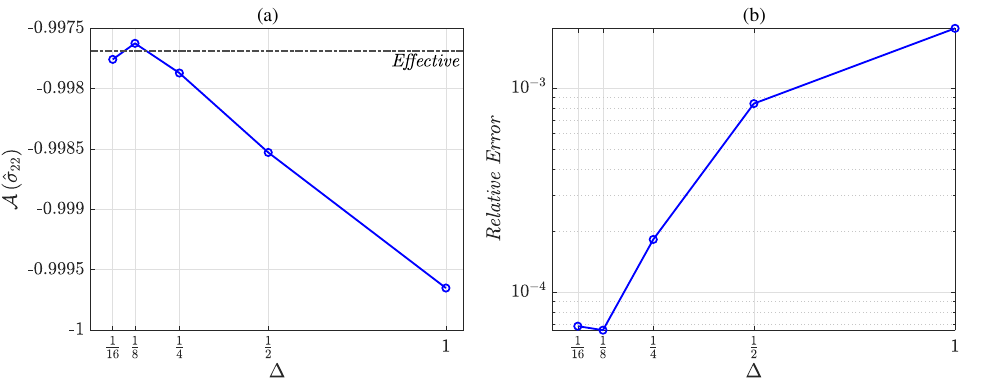}
\caption{{(a) Averaged $\hat{\sigma}_{22}$ at $w = 0.505$  (where $w$ is the transform variable) {in the
    calendering problem} compared to the effective result, and (b)
  {the corresponding relative error as functions of $\Delta$.}}}
\label{calendering2}
\end{figure}

\begin{figure}
	\centering
	\includegraphics[scale=0.95]{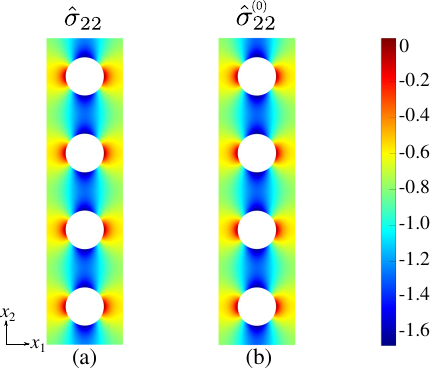}
	\caption{{(a) Prediction of $\hat{\sigma}_{22}$ in the calendaring problem by direct numerical simulation of the full model, and (b) the micro-resolved leading order stress $\hat{\sigma}_{22}^{(0)}$ given by~\eqref{laughinggas}. Results are shown for $\Delta = 1/4$ and $w = 0.505$ (where $w$ is the transform variable).}}
	\label{laughinggasfig}
\end{figure}

\begin{figure}
	\centering
	\includegraphics[scale=0.6]{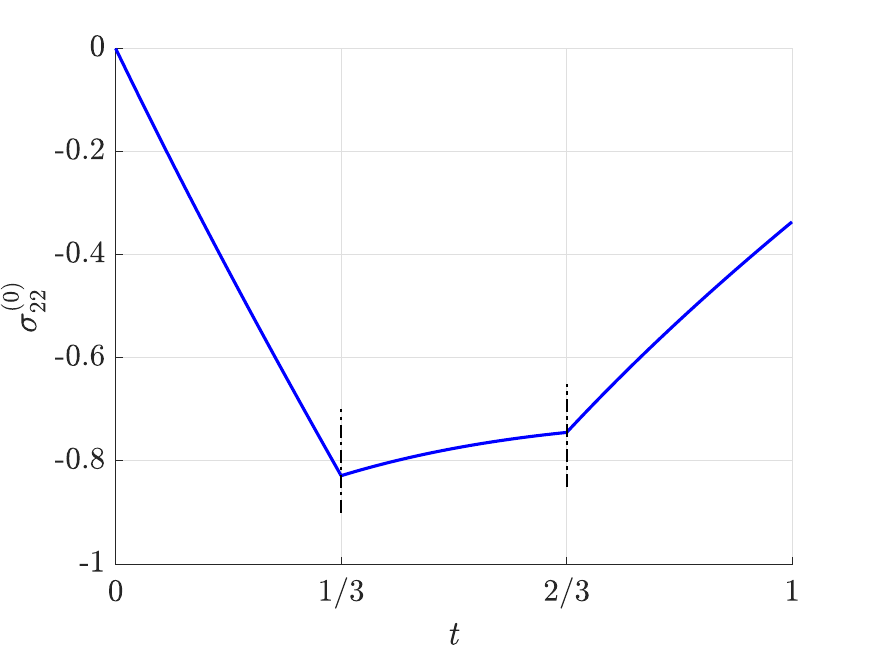}
	\caption{{The time-dependent response of the of the stress ${\sigma}_{22}^{(0)}\left(x_1,x_2,t\right)$ for the calendaring problem in the case $\Delta = 1/4$ at the four locations $x_1 = -0.375$ and $x_2 =\left\{ 0.125,0.375,0.625,0.875 \right\}$. Note that the response at these four locations is identical owing to the symmetry of the unit cell.}}
	\label{mrweewee}
\end{figure}

\subsection{Cell construction and cycling} 
After calendering, electrodes are constructed as part of operational
cells. A separator is sandwiched between and anode and cathode
(negative and positive electrode respectively), and the assembly is then
soaked in electrolyte, which causes the binder to swell. Subsequently,
electrochemical cycling occurs, causing volumetric changes in the
active material particles. We apply the following boundary conditions 
\be
\label{BCbot} \ueff_1 = \ueff_2 = \frac{\dd \peff}{\dd x_2} = 0 \quad \text{on} \quad x_2 = 0, \quad x_2 = 1,
\ee
corresponding to the current collectors being immobile and
impermeable. Mechanical and geometrical parameters are summarised in
Table~\ref{Tab1} and we take 
{\begin{equation}\label{liner}
\begin{array}{r}
{\mathcal G} = 0.5, \quad g(t) = \left\{\begin{array}{r}
 {\mathcal H}(t-2\pi) \sin(t) \quad \text{on} \quad 0 \leq x_2 \leq 1/2, \\[0.1cm]
 -{\mathcal H}(t-2\pi) \sin(t) \quad \text{on} \quad 1/2 < x_2 \leq 1, \\*[2mm]
\end{array}\right. \\  
{\mathcal B} = 5, \quad \beta(t) ={\mathcal H}(t-1),
\end{array}
\end{equation}}
so that the binder swells at $t=1$ and cyclic cell operation (causing
particle expansion in the anode/cathode) occurs for
$t>2\pi$. Parameter values were chosen to model realistic amounts of
binder swelling (for PVDF) and for graphitic particle expansion. The
timescales for the swelling of the binder and particles are
$O(1\text{hr})$ which is markedly longer than that needed to dissipate
pressure gradients by flow of the electrolyte through the pore
space. Hence,
${\mathcal P} \ll 1$ and so it is appropriate to exploit the
asymptotic limit ${\mathcal P} \to 0$, in which $\peff = 0$. We shall
assume stress- and strain-free initial state (\ref{usually}). 

{The deformation is one-dimensional (purely in the $x_2$-direction) and depends only upon the vertical position. Thus
\be
\ueff_1 = \epsiloneff_{11} = \epsiloneff_{12} = \sigmaeff_{11} = \sigmaeff_{12} = 0.\label{BCbot0}
\ee 
The Laplace transform of the non-zero effective deformation, strain and stress components are given by
\be
\label{result234} \ds \hat{\sigma}_{22}^{\text{eff}} = \int_0^1 \lambda(x_2) \,  dx_2, \quad   \hat{\epsilon}_{22}^{\text{eff}} = \frac{ \displaystyle\int_0^1 \lambda(x_2) \,  dx_2 - \lambda(x_2)}{\hat{s}_{11}^{11,\text{eff}}+S^{11,\text{eff}}}, \\*[2mm] \label{result235} \hat{u}_{2}^{\text{eff}} = \frac{x_2 \displaystyle\int_0^1 \lambda(x_2) \,  dx_2 - \int_0^{x_2} \lambda(x_2)}{\hat{s}_{11}^{11,\text{eff}}+S^{11,\text{eff}}} \\*[2mm] \text{where} \quad \lambda(x_2) = \hat{S}^{\beta,\text{eff}}{\mathcal B}\left( K_2 K_\tau w \hat{\beta}+K_1 \hat{\beta} \right) + \hat{S}^{g,\text{eff}} {\mathcal G} \hat{g}.
\ee}
These are compared to an analogous solution of the full model in
Fig.~\ref{pouch1} for various values of $\Delta$. {In addition to
  \eqref{liner}, the boundary conditions to be applied in the
  simulations are 
\be
u_i = 0 &\quad& \text{on} \quad x_2 = 0\quad  \text{ and }\quad  x_2 = 1, \\
u_i,& \quad& \text{and} \quad \sigma_{ij}n_j \quad \text{are periodic in
  $x_1$ with period $\Delta$.}   
\ee}
The expected convergence is observed. We see that the whole cell is under compression owing to net expansion of the internal components.

{The Laplace transform of the micro-resolved stress can be obtain by leveraging (\ref{totals}) along with (\ref{result234})-(\ref{result235}), which yield
\be \label{flatcurve}
\hat{\sigma}_{22}^{(0)} = \left( \hat{s}_{22}^{22} + \hat{S}^{22} \right) \hat{\epsilon}_{22}^{\text{eff}} + \hat{S}^{\beta}{\mathcal B}\left( K_2 K_\tau w \hat{\beta}+K_1 \hat{\beta} \right) + \left( \hat{S}^{g} + \hat{s}_{22}^{g}\right) {\mathcal G} \hat{g}.
\ee
Notably, although $\hat{s}_{22}^{g,\text{eff}}$ does not appear in (\ref{result234}), we must include $\hat{s}_{22}^{g}$ in (\ref{flatcurve}). This arises because even though $\mathcal{A}\left( \hat{s}_{ij}^{g} \right)=0$ the stress $\hat{s}_{ij}^{g}$ is non-zero within the unit cell. The result (\ref{flatcurve}) is compared to the 22-component of the stress obtained by direct numerical simulation of the full model in Fig.~\ref{qotsa}. Good agreement is observed. Both the micro-resolved and direct numerical results concur that the upper- and lower-most particles are displaced upwards by 0.00070 whilst those in the middle are displaced upwards by 0.0021. The upward motion of the particles occurs because the upper electrode is less swollen that the lower one; even though the binder has expanded equally in both electrodes the particles in the upper electrode are delithiated (shrunk) whereas those in the lower electrode have lithiated (grown). Thus, the upward motion of the particles acts to relieve the additional in the lower electrode and redistributes the 22-component of the stress equally throughout.}


\begin{figure}
\centering
\includegraphics[scale=0.7]{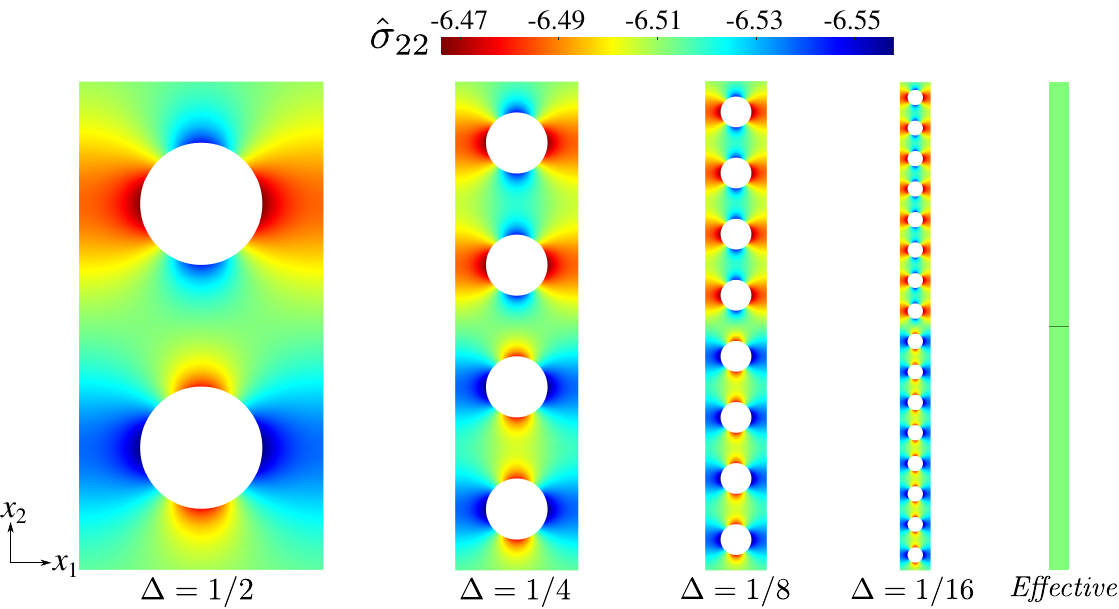}
\caption{{Predictions of $\hat{\sigma}_{22}$ at
    $w = 0.505$ (where $w$ is the transform variable)
    for the pouch cell construction and cycling problem by direct numerical simulations vs. the counterpart effective quantity according to (\ref{result234})}. } 
\label{pouch1}
\end{figure}

\begin{figure}[H]
	\centering
	\includegraphics[scale=0.95]{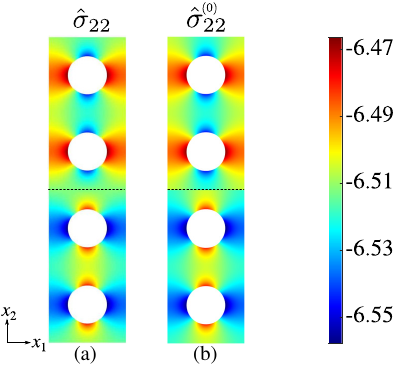}
	\caption{{(a) Prediction of $\hat{\sigma}_{22}$ by direct numerical simulation of the full model for the cell construction and cycling problem, and (b) the micro-resolved leading order stress $\hat{\sigma}_{22}^{(0)}$ given by~\eqref{flatcurve}. Results are shown for $\Delta = 1/4$ and $w = 0.505$ (where $w$ is the transform variable).}}
	\label{qotsa}
\end{figure}

\subsection{Cell impact}
A significant challenge in lithium-ion batteries is designing cells
which are able to withstand impact, or fail, in such a way that
catastrophic thermal runaway or explosion is avoided. Understanding
the stress and deformation within the electrodes themselves is key
because they are the main reservoirs for the explosive components: the
lithium. It is therefore relevant to examine the reaction of an
electrode when it is subjected to rapid loading, as might be
experienced in an electric vehicle during a road traffic accident. As
before we shall assume that the electrode is bonded to a rigid
impermeable current collector on its lower surface whilst a purely
normal load is applied to the upper surface where the electrolyte is
free to enter/leave at constant pressure via the highly flexible
separator. {Thus, we} impose that   
\be
\label{BCbot2} \ueff_1 = \ueff_2 = \frac{\dd \peff}{\dd x_2} = 0 \quad& \text{on} &\quad x_2 = 0,\\
\label{BCtop3} \sigmaeff_{12} = \peff = 0, \quad \sigmaeff_{22}=\Sigma(t), \quad &\text{on}& \quad x_2 = 1,\qquad
\text{where} \quad \Sigma(t) = 0.25\mathcal{H}(t).
\ee
We shall assume a stress- and strain-free initial state, such that appropriate initial conditions are given by (\ref{usually}). In the interest of isolating the effects of the impact we shall assume that the binder and electrode are both unswollen and will therefore select
\be
{\mathcal G} = {\mathcal B} = 0, \quad {\mathcal P}=1.
\ee
The latter is chosen for convenience and corresponds to the scenario in which the timescales of impact, and that of liquid flow, coincide. The remaining geometrical and material parameters are given in Table~\ref{Tab1}.

Integrating each of the equations in (\ref{snorton}) and applying (\ref{BCbot2}b,c) and (\ref{BCtop3}a,b) yields
\be \label{knobsally}
\sigmaeff_{12} =0 , \qquad \sigmaeff_{22} - p = \Sigma(t), \qquad \kappa^{\text{eff}}_{22} \frac{\dd \peff}{\dd x_2} - {\phi}\frac{\dd \ueff_2}{\dd t} = 0.
\ee
Summing the 22-components of (\ref{4eff}) and (\ref{twiglet}) and taking a Laplace transform  yields 
\be \label{flump}
\hat{\sigma}^{\text{eff}}_{22} = \left( \hat{s}^{11,\text{eff}}_{11} + \hat{S}^{11,\text{eff}}\right) \frac{\partial \hat{u}^{\text{eff}}_2}{\partial x_2} 
\ee
Taking a derivative of (\ref{knobsally}c) with respect to $x_2$, taking a Laplace transform of the result, and then eliminating the transform of the pressure using (\ref{knobsally}b) and the transform of the deformation gradient using (\ref{flump}), leads to the following second order ODE 
\be
\kappa^{\text{eff}}_{22} \frac{\dd^2 \hat{\sigma}^{\text{eff}}_{22}}{\dd x_2^2} = \frac{ \hat{\sigma}^{\text{eff}}_{22} w{\phi}}{\hat{s}^{11,\text{eff}}_{11} + \hat{S}^{11,\text{eff}} },
\ee
boundary conditions for which are
\be
\frac{\dd \hat{\sigma}^{\text{eff}}_{22}}{\dd x_2}=0 \quad \text{on} \quad x_2= 0, \qquad \hat{\sigma}^{\text{eff}}_{22} = \hat{\Sigma} \quad \text{on} \quad x_2=1 ,
\ee
where the former arises from (\ref{BCbot2}c) and (\ref{knobsally}b). The solution to this problem is
\be
\hat{\sigma}^{\text{eff}}_{22} = \hat{\Sigma} \frac{\cosh \left( \sqrt{\lambda(w)} x_2 \right)}{\cosh \left( \sqrt{\lambda(w)} \right)} \quad \text{where} \quad \lambda(w) = \frac{w{\phi}}{\kappa^{\text{eff}}_{22} \left( \hat{s}^{11,\text{eff}}_{11} + \hat{S}^{11,\text{eff}} \right) }. \label{talentisalie}
\ee
This is compared to analogous solutions of the full model in Fig.~\ref{Impact}. {In addition to ${\mathcal G} = {\mathcal B} = 0$, the boundary conditions to be applied in the simulations are
\be
u_i = 0 &\text{on}& x_2 = 0, \\
\sigma_{22} = \Sigma(t), \quad \text{and} \quad p = 0 &\text{on}& x_2 = 1, \\
u_i, \quad \text{and} \quad \sigma_{ij}n_j \quad &&\text{are periodic in
  $x_1$ with period $\Delta$.}   
\ee}
The rapid application of tension to the solid skeleton causes a pressure gradient to be established which is slowly dissipated as electrolyte enters the electrode from above.

{We obtain the micro-resolved quantities by substituting (\ref{talentisalie}a) into (\ref{flump}) to obtain an expression for ${\partial \hat{u}^{\text{eff}}_2}/{\partial x_2}$. In turn this can be substituted into (\ref{totals}) to obtain the leading order micro-resolved stress as
\be
\hat{\sigma}_{22}^{(0)} = \left( \hat{s}_{22}^{22} + \hat{S}^{22}\right)\frac{\hat{\Sigma}}{\hat{s}^{11,\text{eff}}_{11} + \hat{S}^{11,\text{eff}}}\frac{\cosh \left( \sqrt{\lambda(w)} x_2 \right)}{\cosh \left( \sqrt{\lambda(w)} \right)}.\label{nomono}
\ee
{The micro-resolved flow can be obtained by returning to (\ref{Darcy}), applying the scalings (\ref{scalings})-(\ref{sc}), and dropping stars yielding $\phi_f (w_i - \dd u_i / \dd t) = - \dd p / \dd x_i$. Applying the multiple scales assumption, namely (\ref{snowdog}), and expanding the dependent variable in a asymptotic series in $\Delta$ following (\ref{lo}) yields a hierarchy of equations that, at leading order, imply that the fluid pressure is independent of microscopic position (in agreement with (\ref{lop})). At next order we find that
	\be
	\phi_f \left( w_i^{(0)} - \frac{\dd u^{(0)}_i}{\dd t} \right) = - \left( \frac{\dd p^{(0)}}{\dd x_i} + \frac{\dd p^{(1)}}{\dd X_i} \right)
	\ee
	Leveraging (\ref{press}) to eliminate $p^{(1)}$ yields
	\be \label{jazzmusic}
	\phi_f \left( w_i^{(0)} - \frac{\dd u^{(0)}_i}{\dd t} \right) = - \left( \delta_{iq} + \frac{\dd p^{q}}{\dd X_i} \right) \frac{\dd p^{(0)}}{\dd x_q}.
	\ee
	In the present problem, for cell impact, we can obtain an expression for the macroscopic gradient in the leading order pressure by substituting (\ref{nomono}) into (\ref{snorton}b), and setting $\mathcal{P}=1$ yielding
	\be \label{diddlydiddy}
	\frac{\partial\hat{p}^{(0)}}{\partial x_1} = 0 , \quad \frac{\partial\hat{p}^{(0)}}{\partial x_2} =  \frac{ \hat{\Sigma}\sqrt{\lambda(w)}}{\cosh \left( \sqrt{\lambda(w)} \right)}\sinh \left( \sqrt{\lambda(w)} x_2 \right).
	\ee
	Taking a Laplace transform of (\ref{jazzmusic}) and eliminating $\dd p^{(0)}/{\dd x_q}$ from the result using (\ref{diddlydiddy}) gives us an expression for the leading order transformed volume-averaged relative fluid velocity.} A comparison of these results against those obtained by direct numerical simulation of the full model is shown in Fig.~\ref{torch}. Excellent agreement is observed in terms of both the stresses and flow.}

\begin{figure}
\centering
\includegraphics[scale=0.6]{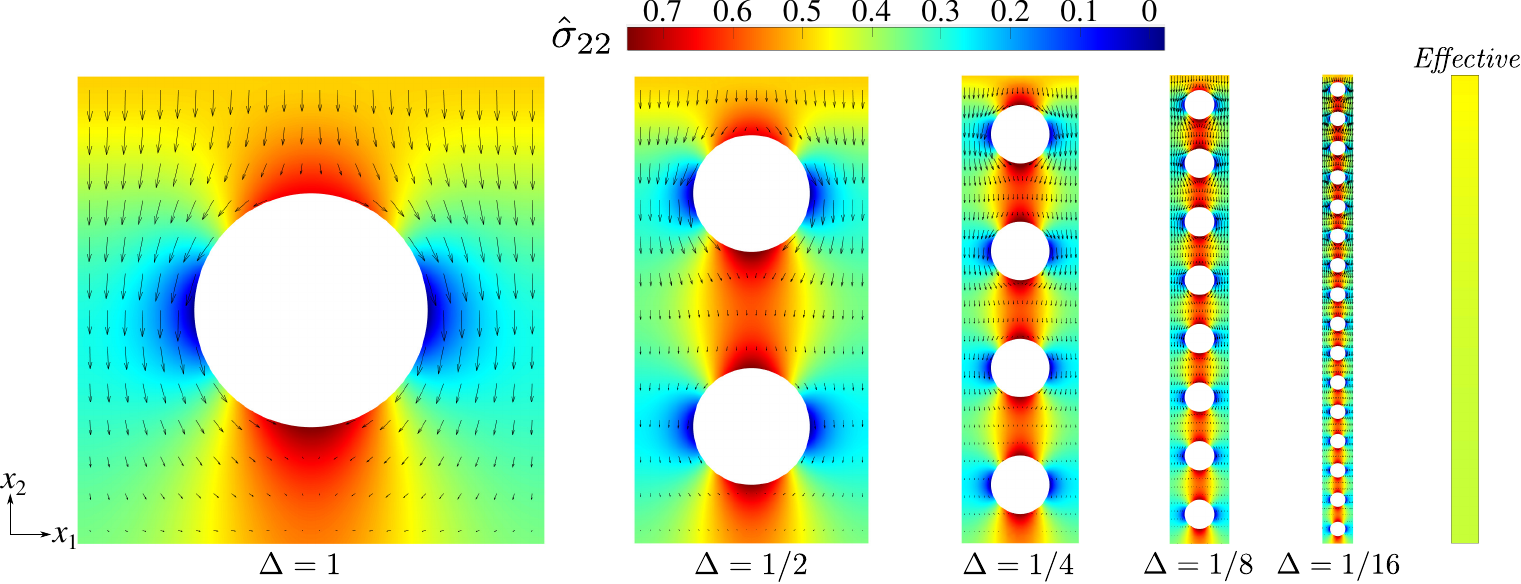}
\caption{{Predictions of $\hat{\sigma}_{22}$ at
  $w = 0.505$ (where $w$ is the transform variable)
  for the pouch cell impact problem obtained by direct numerical simulation of the full model along with the solution of the
  homogenised equations, (\ref{talentisalie}). Arrows indicate the vector field of $-\dd \hat{p}/\dd x_i$, or equivalently {the transform of the volume-averaged relative fluid velocity, $\phi_f (w_i - \dd u_i/\dd t)$}.}}
\label{Impact}
\end{figure}

\begin{figure}
	\centering
	\includegraphics[scale=0.95]{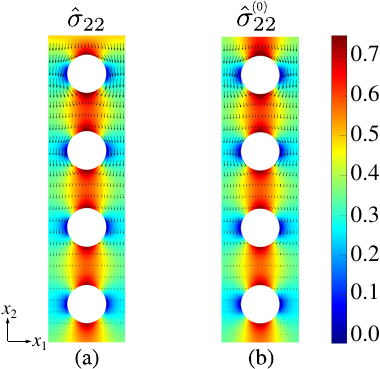}
	\caption{{(a) Prediction of $\hat{\sigma}_{22}$ and {$-\dd \hat{p}/\dd x_i$} by direct numerical simulation of the full model for the cell impact problem, and (b) the micro-resolved leading order 22-component of the stress and {the transform of the volume-averaged relative fluid velocity}, given by equations \eqref{nomono} and \eqref{jazzmusic}-\eqref{diddlydiddy} respectively. Here $\Delta = 1/4$ and $w = 0.505$ (where $w$ is the transform variable).}}
	\label{torch}
\end{figure}

\section{\label{sec7}Conclusions}
We have presented a model for the mechanics of composites where rigid
inclusions are embedded within a porous viscoelastic material whose
pores contain fluid. The application that originally motivated this
work is the porous electrodes found in modern lithium-ion batteries,
but the results of our analysis could be applied elsewhere, e.g. to
concrete. The disparity in lengthscales between that of a single
inclusion and that of the composite as a whole has been exploited to
carry out a systematic multiple scale homogenisation, yielding an
effective system of equations that accurately describes the mechanical
behaviour at the macroscopic level (that of the composite as a
whole). The effective model is similar to form to the original
microscopic model, but the effective constitutive behaviour exhibits
more relaxation timescales than the constituent viscoelastic, and
sources of volumetric stress are introduced to account for the
swelling/contract of the phases at the microscopic scale. The expected
convergence of the effective model to the original system has been
demonstrated {computationally} in the limit that this ratio of
lengthscales tends to zero. Three relevant scenarios have been
simulated, namely: (i) electrode calendering where the composite is
compressed under a steel roller, (ii) electrode construction and
(dis)charge where the electrode is wetted with liquid electrolyte
(causing the viscoelastic polymer to swell) and the inclusions
(electrode particles) expand/contract, and (iii) cell impact, where
the composite is subjected to a sudden impact representative of an
electric vehicle crash. 

The effective equations are markedly {simpler and hence} cheaper to solve computationally than the original model, and this opens the door to carrying out mechanical simulations of entire batteries; this certainly would not have been possible via direct computation of the original model posed on the original geometry, even with modern computing resources. The value of this to the lithium-ion battery industry is substantial. {Importantly, the micro-resolved stress, strain, displacement and pressure fields are not lost during the homogenisation, and can be recovered having solved the effective equations. Thus, microscopic effects, such degradation via de-bonding of the binder from the particle surfaces, can be meaningfully investigated.} The effective model can be used to identify strategies for mitigating the mechanical damage that occurs in real devices, contributing to their limited lifetime. An extension to the service life of lithium-ion batteries gives rises to proportionately reduced cost of manufacture and this is sorely needed in the burgeoning electric vehicle market.

Whilst this work is an important contribution to realising truly useful battery simulation tools, there are a number of deficiencies that should be address in future work. First, contact between the particles has been precluded, but almost certainly occurs where strains are moderately large. Particle-particle contact is likely to be important because it may be a contributor to electrode particle fracture and this is known to be a catalyst for various forms of chemical degradation which accelerate cell aging. Second, the microstructures of real electrodes are not as regular as that which has been used here. It would be informative to see how the rheology of the effective material is altered by the microscale geometry, and in particular when the particles are allowed to have varying size, shape and position (as is the case in reality). Third, some model for debonding between the polymer and electrode particles should be considered. Here, we have assumed that the two phases remain adhered to one another, but this is unlikely to be true through a battery's lifetime. The loss of contact will undoubtedly affect the behaviour of the composite but is also important to detect from a practical standpoint because it signals the loss of electrical contact. Isolated electrode particles are electrochemically inert, and therefore no longer contribute to cell capacity.

\section*{Acknowledgements}
JF and SJC were supported by the Faraday Institution Multi-Scale Modelling (MSM) project (grant number EP/S003053/1). JF and AG were supported by the Engineering and Physical Sciences Research Council (grant number EP/T000775/1).


\begin{thebibliography}{10}

\bibitem{Kan06}
K.~Kang, Y.~S. Meng, J.~Br{\'e}ger, C.~P. Grey, and G.~Ceder, ``Electrodes with
  high power and high capacity for rechargeable lithium batteries,'' {\em
  Science}, vol.~311, no.~5763, pp.~977--980, 2006.

\bibitem{Zha19}
Y.~Zhao, P.~Stein, Y.~Bai, M.~Al-Siraj, Y.~Yang, and B.-X. Xu, ``A review on
  modeling of electro-chemo-mechanics in lithium-ion batteries,'' {\em Journal
  of Power Sources}, vol.~413, pp.~259--283, 2019.

\bibitem{Owe97}
J.~R. Owen, ``Rechargeable lithium batteries,'' {\em Chemical Society Reviews},
  vol.~26, no.~4, pp.~259--267, 1997.

\bibitem{Vet05}
J.~Vetter, P.~Nov{\'a}k, M.~R. Wagner, C.~Veit, K.-C. M{\"o}ller, J.~Besenhard,
  M.~Winter, M.~Wohlfahrt-Mehrens, C.~Vogler, and A.~Hammouche, ``Ageing
  mechanisms in lithium-ion batteries,'' {\em Journal of power sources},
  vol.~147, no.~1-2, pp.~269--281, 2005.

\bibitem{Eta11}
V.~Etacheri, R.~Marom, R.~Elazari, G.~Salitra, and D.~Aurbach, ``Challenges in
  the development of advanced li-ion batteries: a review,'' {\em Energy \&
  Environmental Science}, vol.~4, no.~9, pp.~3243--3262, 2011.

\bibitem{Saf09}
M.~Safari, M.~Morcrette, A.~Teyssot, and C.~Delacourt, ``Multimodal
  physics-based aging model for life prediction of li-ion batteries,'' {\em
  Journal of The Electrochemical Society}, vol.~156, no.~3, pp.~A145--A153,
  2009.

\bibitem{Bir17}
C.~R. Birkl, M.~R. Roberts, E.~McTurk, P.~G. Bruce, and D.~A. Howey,
  ``Degradation diagnostics for lithium ion cells,'' {\em Journal of Power
  Sources}, vol.~341, pp.~373--386, 2017.

\bibitem{McD16}
M.~T. McDowell, S.~Xia, and T.~Zhu, ``The mechanics of large-volume-change
  transformations in high-capacity battery materials,'' {\em Extreme Mechanics
  Letters}, vol.~9, pp.~480--494, 2016.

\bibitem{Len17}
G.~Lenze, F.~R{\"o}der, H.~Bockholt, W.~Haselrieder, A.~Kwade, and U.~Krewer,
  ``Simulation-supported analysis of calendering impacts on the performance of
  lithium-ion-batteries,'' {\em Journal of The Electrochemical Society},
  vol.~164, no.~6, pp.~A1223--A1233, 2017.

\bibitem{Kwa18}
A.~Kwade, W.~Haselrieder, R.~Leithoff, A.~Modlinger, F.~Dietrich, and
  K.~Droeder, ``Current status and challenges for automotive battery production
  technologies,'' {\em Nature Energy}, vol.~3, no.~4, pp.~290--300, 2018.

\bibitem{Kes15}
M.~Kespe and H.~Nirschl, ``Numerical simulation of lithium-ion battery
  performance considering electrode microstructure,'' {\em International
  Journal of Energy Research}, vol.~39, no.~15, pp.~2062--2074, 2015.

\bibitem{Wan04}
C.-W. Wang, Y.-B. Yi, A.~Sastry, J.~Shim, and K.~Striebel, ``Particle
  compression and conductivity in li-ion anodes with graphite additives,'' {\em
  Journal of the Electrochemical Society}, vol.~151, no.~9, pp.~A1489--A1498,
  2004.

\bibitem{Gim19}
C.~S. Gim{\'e}nez, B.~Finke, C.~Schilde, L.~Frob{\"o}se, and A.~Kwade,
  ``Numerical simulation of the behavior of lithium-ion battery electrodes
  during the calendaring process via the discrete element method,'' {\em Powder
  technology}, vol.~349, pp.~1--11, 2019.

\bibitem{Gim20}
C.~S. Gim{\'e}nez, L.~Helmers, C.~Schilde, A.~Diener, and A.~Kwade, ``Modeling
  the electrical conductive paths within all-solid-state battery electrodes,''
  {\em Chemical Engineering \& Technology}, 2020.

\bibitem{Ant15}
D.~Antartis, S.~Dillon, and I.~Chasiotis, ``Effect of porosity on
  electrochemical and mechanical properties of composite li-ion anodes,'' {\em
  Journal of Composite Materials}, vol.~49, no.~15, pp.~1849--1862, 2015.

\bibitem{Zhe12}
H.~Zheng, L.~Tan, G.~Liu, X.~Song, and V.~S. Battaglia, ``Calendering effects
  on the physical and electrochemical properties of li [ni1/3mn1/3co1/3] o2
  cathode,'' {\em Journal of Power Sources}, vol.~208, pp.~52--57, 2012.

\bibitem{Has13}
W.~Haselrieder, S.~Ivanov, D.~K. Christen, H.~Bockholt, and A.~Kwade, ``Impact
  of the calendering process on the interfacial structure and the related
  electrochemical performance of secondary lithium-ion batteries,'' {\em ECS
  Transactions}, vol.~50, no.~26, p.~59, 2013.

\bibitem{Che06}
L.~Chen, X.~Xie, J.~Xie, K.~Wang, and J.~Yang, ``Binder effect on cycling
  performance of silicon/carbon composite anodes for lithium ion batteries,''
  {\em Journal of applied electrochemistry}, vol.~36, no.~10, pp.~1099--1104,
  2006.

\bibitem{Cho14}
S.-L. Chou, Y.~Pan, J.-Z. Wang, H.-K. Liu, and S.-X. Dou, ``Small things make a
  big difference: binder effects on the performance of li and na batteries,''
  {\em Physical Chemistry Chemical Physics}, vol.~16, no.~38, pp.~20347--20359,
  2014.

\bibitem{Mag10}
A.~Magasinski, B.~Zdyrko, I.~Kovalenko, B.~Hertzberg, R.~Burtovyy, C.~F.
  Huebner, T.~F. Fuller, I.~Luzinov, and G.~Yushin, ``Toward efficient binders
  for li-ion battery si-based anodes: polyacrylic acid,'' {\em ACS applied
  materials \& interfaces}, vol.~2, no.~11, pp.~3004--3010, 2010.

\bibitem{Kim04}
I.-S. Kim and P.~N. Kumta, ``High capacity si/c nanocomposite anodes for li-ion
  batteries,'' {\em Journal of Power Sources}, vol.~136, no.~1, pp.~145--149,
  2004.

\bibitem{Xia11}
X.~Xiao, P.~Liu, M.~Verbrugge, H.~Haftbaradaran, and H.~Gao, ``Improved cycling
  stability of silicon thin film electrodes through patterning for high energy
  density lithium batteries,'' {\em Journal of Power Sources}, vol.~196, no.~3,
  pp.~1409--1416, 2011.

\bibitem{Ebn13}
M.~Ebner, F.~Marone, M.~Stampanoni, and V.~Wood, ``Visualization and
  quantification of electrochemical and mechanical degradation in li ion
  batteries,'' {\em Science}, vol.~342, no.~6159, pp.~716--720, 2013.

\bibitem{Liu12}
G.~Liu, H.~Zheng, X.~Song, and V.~S. Battaglia, ``Particles and polymer binder
  interaction: a controlling factor in lithium-ion electrode performance,''
  {\em Journal of The Electrochemical Society}, vol.~159, no.~3,
  pp.~A214--A221, 2012.

\bibitem{Kov11}
I.~Kovalenko, B.~Zdyrko, A.~Magasinski, B.~Hertzberg, Z.~Milicev, R.~Burtovyy,
  I.~Luzinov, and G.~Yushin, ``A major constituent of brown algae for use in
  high-capacity li-ion batteries,'' {\em Science}, vol.~334, no.~6052,
  pp.~75--79, 2011.

\bibitem{Shi02}
J.~Shim, R.~Kostecki, T.~Richardson, X.~Song, and K.~A. Striebel,
  ``Electrochemical analysis for cycle performance and capacity fading of a
  lithium-ion battery cycled at elevated temperature,'' {\em Journal of power
  sources}, vol.~112, no.~1, pp.~222--230, 2002.

\bibitem{Che13}
J.~Chen, J.~Liu, Y.~Qi, T.~Sun, and X.~Li, ``Unveiling the roles of binder in
  the mechanical integrity of electrodes for lithium-ion batteries,'' {\em
  Journal of The Electrochemical Society}, vol.~160, no.~9, pp.~A1502--A1509,
  2013.

\bibitem{San09}
S.~Santhanagopalan, P.~Ramadass, and J.~Z. Zhang, ``Analysis of internal
  short-circuit in a lithium ion cell,'' {\em Journal of Power Sources},
  vol.~194, no.~1, pp.~550--557, 2009.

\bibitem{Cai11}
W.~Cai, H.~Wang, H.~Maleki, J.~Howard, and E.~Lara-Curzio, ``Experimental
  simulation of internal short circuit in li-ion and li-ion-polymer cells,''
  {\em Journal of Power Sources}, vol.~196, no.~18, pp.~7779--7783, 2011.

\bibitem{Chr06}
J.~Christensen and J.~Newman, ``A mathematical model of stress generation and
  fracture in lithium manganese oxide,'' {\em Journal of The Electrochemical
  Society}, vol.~153, no.~6, pp.~A1019--A1030, 2006.

\bibitem{Chr06b}
J.~Christensen and J.~Newman, ``Stress generation and fracture in lithium
  insertion materials,'' {\em Journal of Solid State Electrochemistry},
  vol.~10, no.~5, pp.~293--319, 2006.

\bibitem{Zha07}
X.~Zhang, W.~Shyy, and A.~M. Sastry, ``Numerical simulation of
  intercalation-induced stress in li-ion battery electrode particles,'' {\em
  Journal of the Electrochemical Society}, vol.~154, no.~10, pp.~A910--A916,
  2007.

\bibitem{Kli16}
M.~Klinsmann, D.~Rosato, M.~Kamlah, and R.~M. McMeeking, ``Modeling crack
  growth during li insertion in storage particles using a fracture phase field
  approach,'' {\em Journal of the Mechanics and Physics of Solids}, vol.~92,
  pp.~313--344, 2016.

\bibitem{Ai20}
W.~Ai, L.~Kraft, J.~Sturm, A.~Jossen, and B.~Wu, ``Electrochemical
  thermal-mechanical modelling of stress inhomogeneity in lithium-ion pouch
  cells,'' {\em Journal of The Electrochemical Society}, vol.~167, no.~1,
  p.~013512, 2020.

\bibitem{Rie16}
B.~Rieger, S.~V. Erhard, K.~Rumpf, and A.~Jossen, ``A new method to model the
  thickness change of a commercial pouch cell during discharge,'' {\em Journal
  of The Electrochemical Society}, vol.~163, no.~8, pp.~A1566--A1575, 2016.

\bibitem{Timms2022}
R.~Timms, S.~Psaltis, C.~P. Please, and S.~J. Chapman, ``Asymptotic reduction
  of a mechanical model for buckling behaviour in cylindrical batteries,''
  preprint.

\bibitem{Oh16}
K.-Y. Oh, B.~I. Epureanu, J.~B. Siegel, and A.~G. Stefanopoulou,
  ``Phenomenological force and swelling models for rechargeable lithium-ion
  battery cells,'' {\em Journal of Power Sources}, vol.~310, pp.~118--129,
  2016.

\bibitem{Dai17}
H.~Dai, C.~Yu, X.~Wei, and Z.~Sun, ``State of charge estimation for lithium-ion
  pouch batteries based on stress measurement,'' {\em Energy}, vol.~129,
  pp.~16--27, 2017.

\bibitem{Wu14}
W.~Wu, X.~Xiao, M.~Wang, and X.~Huang, ``A microstructural resolved model for
  the stress analysis of lithium-ion batteries,'' {\em Journal of The
  Electrochemical Society}, vol.~161, no.~5, pp.~A803--A813, 2014.

\bibitem{Kan09}
P.~Kanout{\'e}, D.~Boso, J.~Chaboche, and B.~Schrefler, ``Multiscale methods
  for composites: a review,'' {\em Archives of Computational Methods in
  Engineering}, vol.~16, no.~1, pp.~31--75, 2009.

\bibitem{Fos17}
J.~M. Foster, S.~J. Chapman, G.~Richardson, and B.~Protas, ``A mathematical
  model for mechanically-induced deterioration of the binder in lithium-ion
  electrodes,'' {\em SIAM Journal on Applied Mathematics}, vol.~77, no.~6,
  pp.~2172--2198, 2017.

\bibitem{Fos17b}
J.~M. Foster, X.~Huang, M.~Jiang, S.~J. Chapman, B.~Protas, and G.~Richardson,
  ``Causes of binder damage in porous battery electrodes and strategies to
  prevent it,'' {\em Journal of Power Sources}, vol.~350, pp.~140--151, 2017.

\bibitem{Che04}
Z.~Chen, L.~Christensen, and J.~Dahn, ``Mechanical and electrical properties of
  poly (vinylidene fluoride--tetrafluoroethylene--propylene)/super-s carbon
  black swelled in liquid solvent as an electrode binder for lithium-ion
  batteries,'' {\em Journal of applied polymer science}, vol.~91, no.~5,
  pp.~2958--2965, 2004.

\bibitem{Cherkaev2000}
A.~Cherkaev, {\em Variational Methods for Structural Optimization}, vol.~140 of
  {\em Applied Mathematical Sciences}.
\newblock New York: Springer-Verlag, 2000.

\bibitem{Milton2002}
G.~W. Milton, {\em Theory of Composites}.
\newblock Cambridge: Cambridge University Press, 2002.

\bibitem{Torquato2002}
S.~Torquato, {\em Random Heterogeneous Materials: Microstructure and
  Macroscopic Properties}.
\newblock New York: Springer-Verlag, 2002.

\bibitem{Bur81}
R.~Burridge and J.~B. Keller, ``Poroelasticity equations derived from
  microstructure,'' {\em The Journal of the Acoustical Society of America},
  vol.~70, no.~4, pp.~1140--1146, 1981.

\bibitem{Ogd74}
R.~Ogden, ``On the overall moduli of non-linear elastic composite materials,''
  {\em Journal of the Mechanics and Physics of Solids}, vol.~22, no.~6,
  pp.~541--553, 1974.

\bibitem{Dvo00}
G.~J. Dvorak, ``Composite materials: Inelastic behavior, damage, fatigue and
  fracture,'' {\em International Journal of Solids and Structures}, vol.~37,
  no.~1-2, pp.~155--170, 2000.

\bibitem{Dvo01}
G.~J. Dvorak and J.~Zhang, ``Transformation field analysis of damage evolution
  in composite materials,'' {\em Journal of the Mechanics and Physics of
  Solids}, vol.~49, no.~11, pp.~2517--2541, 2001.

\bibitem{Voy95}
G.~Z. Voyiadjis and T.~Park, ``Local and interfacial damage analysis of metal
  matrix composites,'' {\em International Journal of Engineering Science},
  vol.~33, no.~11, pp.~1595--1621, 1995.

\bibitem{Fis99}
J.~Fish, Q.~Yu, and K.~Shek, ``Computational damage mechanics for composite
  materials based on mathematical homogenization,'' {\em International journal
  for numerical methods in engineering}, vol.~45, no.~11, pp.~1657--1679, 1999.

\bibitem{Bul99}
V.~Bulsara, R.~Talreja, and J.~Qu, ``Damage initiation under transverse loading
  of unidirectional composites with arbitrarily distributed fibers,'' {\em
  Composites science and technology}, vol.~59, no.~5, pp.~673--682, 1999.

\bibitem{Fis01}
J.~Fish and Q.~Yu, ``Multiscale damage modelling for composite materials:
  theory and computational framework,'' {\em International Journal for
  Numerical Methods in Engineering}, vol.~52, no.~1-2, pp.~161--191, 2001.

\bibitem{Seg02}
J.~Segurado and J.~Llorca, ``A numerical approximation to the elastic
  properties of sphere-reinforced composites,'' {\em Journal of the Mechanics
  and Physics of Solids}, vol.~50, no.~10, pp.~2107--2121, 2002.

\bibitem{Dav13}
Y.~Davit, C.~G. Bell, H.~M. Byrne, L.~A. Chapman, L.~S. Kimpton, G.~E. Lang,
  K.~H. Leonard, J.~M. Oliver, N.~C. Pearson, R.~J. Shipley, {\em et~al.},
  ``Homogenization via formal multiscale asymptotics and volume averaging: How
  do the two techniques compare?,'' {\em Advances in Water Resources}, vol.~62,
  pp.~178--206, 2013.

\bibitem{Fey99}
F.~Feyel, ``Multiscale fe2 elastoviscoplastic analysis of composite
  structures,'' {\em Computational Materials Science}, vol.~16, no.~1-4,
  pp.~344--354, 1999.

\bibitem{Hai08}
M.~Hain and P.~Wriggers, ``Numerical homogenization of hardened cement paste,''
  {\em Computational Mechanics}, vol.~42, no.~2, pp.~197--212, 2008.

\bibitem{Con16}
L.~Contrafatto, M.~Cuomo, and S.~Gazzo, ``A concrete homogenisation technique
  at meso-scale level accounting for damaging behaviour of cement paste and
  aggregates,'' {\em Computers \& Structures}, vol.~173, pp.~1--18, 2016.

\bibitem{Hu98}
G.~Hu, G.~Guo, and D.~Baptiste, ``A micromechanical model of influence of
  particle fracture and particle cluster on mechanical properties of metal
  matrix composites,'' {\em Computational Materials Science}, vol.~9, no.~3-4,
  pp.~420--430, 1998.

\bibitem{Par06}
W.~Parnell, Q.~Grimal, I.~Abrahams, and P.~Laugier, ``Modelling cortical bone
  using the method of asymptotic homogenization,'' {\em Journal of
  biomechanics}, no.~39, p.~S20, 2006.

\bibitem{Mul06}
U.~M{\"u}ller, W.~Gindl, and G.~Jeronimidis, ``Biomechanics of a branch--stem
  junction in softwood,'' {\em Trees}, vol.~20, no.~5, pp.~643--648, 2006.

\bibitem{biot}
M.~A. Biot, ``General theory of three-dimensional consolidation,'' {\em Journal
  of applied physics}, vol.~12, no.~2, pp.~155--164, 1941.

\bibitem{bear}
J.~Bear and Y.~Bachmat, {\em Introduction to modeling of transport phenomena in
  porous media}, vol.~4.
\newblock Springer Science \& Business Media, 2012.

\bibitem{wang}
H.~Wang, ``Linear poroelasticity,'' 2000.

\bibitem{Fow97}
A.~C. Fowler, {\em Mathematical models in the applied sciences}, vol.~17.
\newblock Cambridge University Press, 1997.

\bibitem{Cha90}
C.~S. Chang and C.~L. Liao, ``Constitutive relation for a particulate medium
  with the effect of particle rotation,'' {\em International Journal of Solids
  and Structures}, vol.~26, no.~4, pp.~437--453, 1990.

\bibitem{Cha15}
S.~Chapman and S.~Mcburnie, ``Integral constraints in multiple-scales
  problems,'' {\em European Journal of Applied Mathematics}, vol.~26, no.~5,
  pp.~595--614, 2015.

\bibitem{Meyer2017}
C.~Meyer, H.~Bockholt, W.~Haselrieder, and A.~Kwade, ``Characterization of the
  calendering process for compaction of electrodes forlithium-ion batteries,''
  {\em Journal of Materials Processing Technology}, vol.~249, pp.~172--178,
  2017.

\bibitem{Hecht}
F.~Hecht, ``New development in freefem++,'' {\em Journal of Numerical
  Mathematics}, vol.~20(3-4), pp.~251--265, 2012.

\end{thebibliography}

\appendix
\section{Calculating the cell functions}\label{Cellsfunct}
Here we detail the approach for computing the four cell functions $\sigma^{kl}_{ij}$, $\sigma^{\beta}_{ij}$, $\sigma^{g}_{ij}$ and $p^q$ introduced in and around equation (\ref{lo})--(\ref{press}). 

\subsection{\label{macstr}{Response $\sigma_{ij}^{kl}$ to macroscopic deformation gradients}}
Macroscopic deformation gradients force equations
{(\ref{pushed})--(\ref{flap})} in two different locations, namely: in
(\ref{la}) and in (\ref{bridge}). Rather than directly solving for a
cell function that contains both these forcings, we further subdivide
the solution process into two parts. First we characterise the
response of the system to the excitation in (\ref{la}) (see 
\S\ref{train}), and then to the excitation in (\ref{bridge}) (see
\S\ref{macp}). Since the problem is linear, summing the resulting
stress fields yields $\sigma_{ij}^{kl}$. 

\subsubsection{Strain in the composite\label{train}}
The series of cell functions which correspond to the deformations
forced by macroscopic strain (the superscript $kl$ indicating that a macroscopic strain is in the $kl$ component) is 
\be
\label{kick3} \frac{\partial \sigma_{ij}^{kl}}{\partial X_j} =0, \quad \epsilon_{ij}^{kl} = \frac{1}{2} \left( \frac{\partial u^{kl}_i}{\partial X_j} + \frac{\dd u^{kl}_j}{\dd X_i} \right) + \frac{1}{2} \left( \delta_{ik} \delta_{jl} + \delta_{il} \delta_{jk} \right) \delta(t-t'),\\
S^{kl} = \frac{1}{2} \sigma^{kl}_{mm}, \quad E^{kl} = \frac{1}{2} \epsilon^{kl}_{mm},\\
s^{kl}_{ij} = \sigma^{kl}_{ij} - \delta_{ij} S^{kl},\quad e^{kl}_{ij} = \epsilon^{kl}_{ij} - \delta_{ij} E^{kl},\\
{G_{\tau} \dot{s}^{kl}_{ij} + s^{kl}_{ij} = G_2 G_\tau \dot{e}^{kl}_{ij} + G_1 e^{kl}_{ij},} \quad {K_{\tau} \dot{S}^{kl} + S^{kl} = K_2 K_\tau \dot{E}^{kl}  + K_1 E^{kl}}.\label{kick5}
\ee
with the boundary and initial conditions
\be 
u^{kl}_i = U^{kl}_i(t) +\Theta^{kl}(t) \varepsilon_{ij3} X_j \on \Gamma_{inc},\label{kick6}\\
u^{kl}_i \quad \text{and} \quad \sigma_{ij}^{kl} n_j \quad \text{periodic} \on \Gamma_{box},\\
\int_{\Gamma_{inc}} \sigma^{kl}_{ij} {n}_j \, d\Gamma = 0, \qquad \int_{\Gamma_{inc}} \varepsilon_{ik3} \sigma^{kl}_{ij} X_k n_j \, d\Gamma = 0,\\
\label{UUU} \frac{\dd U^{kl}_i}{\dd X_j} = 0, \qquad \frac{\dd \Theta^{kl}}{\dd X_j} = 0,\\
E^{kl}|_{t=0} = 0, \quad S^{kl}|_{t=0} = 0, \quad e_{ij}^{kl}|_{t=0} = 0, \quad s_{ij}^{kl}|_{t=0} = 0. \label{kick4}
\ee
The forcing is provided via the strain equation (\ref{kick3}b) at time $t=t'>0$. In a general two-dimensional RVE there are four such problems to be solved, however, owing to the symmetry of our RVE we can obtain the requisite information by solving two problems {only}. We consider the application of a unit normal strain applied to $\epsilon_{11}$ as well as a unit shear strain applied to $\epsilon_{12}$ and note that the solutions for the excitations applied to $\epsilon_{22}$ and $\epsilon_{21}$ can be obtained by a rotation. We find that 
\be
\label{joe} u^{11}_i = U^{11}_i = \Theta^{11} = 0,\\
\label{11sol1} \epsilon^{11}_{11} = \delta(t-t'), \quad \epsilon^{11}_{12} = \epsilon^{11}_{22} = 0, \\
E^{11} = \frac{\delta(t-t')}{2}, \quad e^{11}_{11} = \frac{\delta(t-t')}{2}, \quad e^{11}_{12} = 0, \\
 \sigma^{11}_{11} = s^{11}_{11} + S^{11}, \quad \sigma^{11}_{12} = 0, \quad \sigma^{11}_{22} = -s^{11}_{11} +S^{11},\\ \label{11sol2} s^{11}_{11} = { \frac{G_2}{2}  \delta(t-t')} + \frac{G_1-G_2}{2G_{\tau}} \exp \left( - \frac{1}{G_\tau} (t-t') \right) {\mathcal H}(t-t'), \quad s^{11}_{12} = 0, \\ \label{11sol3} S^{11} = \frac{K_2}{2} \delta(t-t') + \frac{K_1-K_2}{2K_{\tau}} \exp \left( - \frac{1}{K_\tau} (t-t') \right) {\mathcal H}(t-t'),
\ee
and that
\be
u^{12}_i = U^{12}_i = \Theta^{12} = 0,\\
\label{12sol1} \epsilon^{12}_{11} = 0, \quad \epsilon^{12}_{12} = \frac{\delta(t-t')}{2}, \quad \epsilon^{12}_{22} = 0, \\ 
E^{12} = 0 , \quad e^{12}_{11} = 0, \quad e^{12}_{12} = \frac{\delta(t-t')}{2}, \\ \sigma^{12}_{11} = 0, \quad \sigma^{12}_{12} = s^{12}_{12}, \quad \sigma^{12}_{22} = 0, \\ \label{12sol2} s_{11}^{12}=0, \quad s^{12}_{12} = \frac{G_2}{2} \delta(t-t')+ \frac{G_1-G_2}{2G_{\tau}} \exp \left( - \frac{1}{G_\tau} (t-t') \right) {\mathcal H}(t-t'),\\ \label{satriani} S=0.
\ee
After the application of the Laplace transform, the stress components of this load are shown in Fig.~\ref{MSC}, where the transformed variables are indicated with tilde.

\begin{figure}
\centering
\includegraphics[scale=0.5]{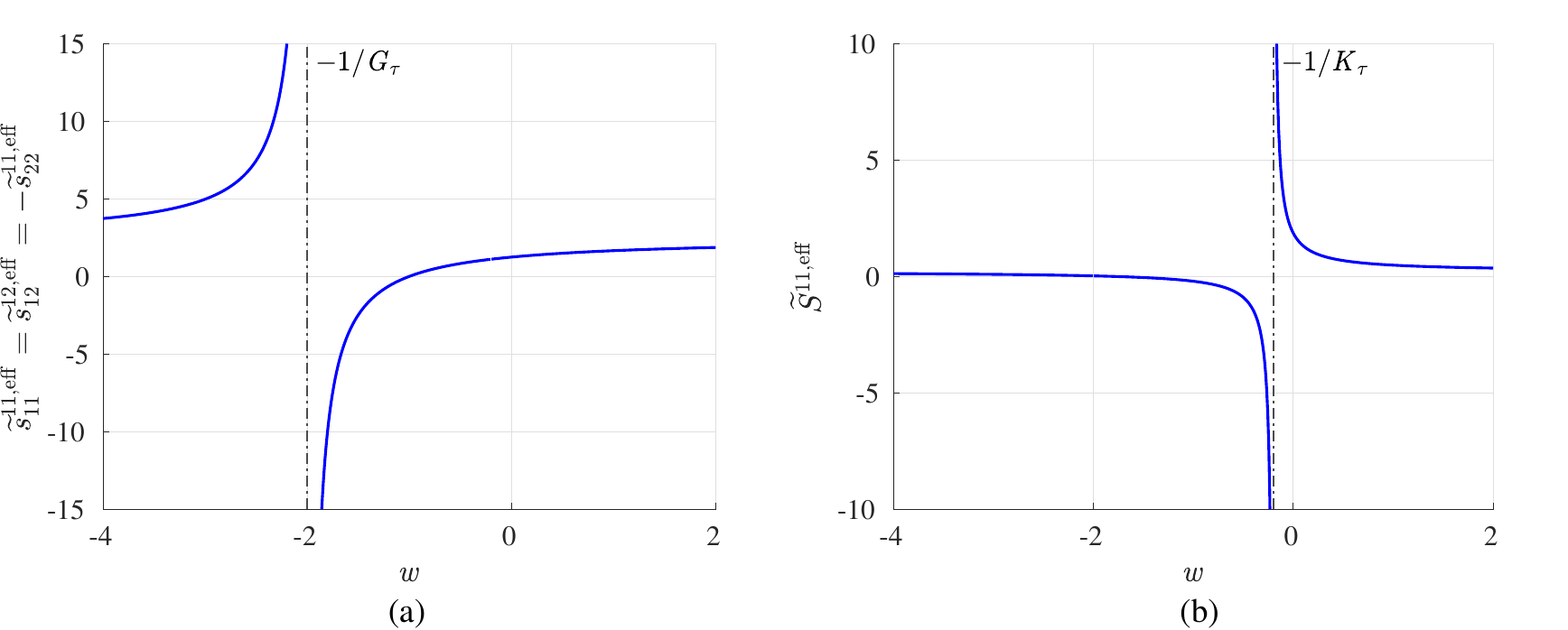}
\caption{The cell functions that characterise the response of the
  effective material due to deformation gradients in the composite
  {obtained as} solutions to
  equations~\eqref{kick3}--\eqref{kick4} after application of the
  averaging operator~\eqref{averagingOp} and Laplace transformation.} 
\label{MSC}
\end{figure}


\subsubsection{\label{macp}Strain between the particles}
{The problems for the cell functions which capture the deformations forced by the macroscopic strain between particles, henceforth denoted with a superscript $mn$, indicating a strain in the $mn$ direction, are given by
\be
\frac{\partial \sigma_{ij}^{mn}}{\partial X_j} =0, \quad \epsilon_{ij}^{mn} = \frac{1}{2} \left( \frac{\partial u^{mn}_i}{\partial X_j} + \frac{\dd u^{mn}_j}{\dd X_i} \right),\label{kate0}\\
S^{mn} = \frac{1}{2} \sigma^{mn}_{kk}, \quad E^{mn} = \frac{1}{2} \epsilon^{mn}_{kk},\\
s^{mn}_{ij} = \sigma^{mn}_{ij} - \delta_{ij} S^{mn},\quad e^{mn}_{ij} = \epsilon^{mn}_{ij} - \delta_{ij} E^{mn},\\
G_{\tau} \dot{s}^{mn}_{ij} + s^{mn}_{ij} = G_2 G_\tau \dot{e}^{mn}_{ij} + G_1 e^{mn}_{ij},} \quad {K_{\tau} \dot{S}^{mn} + S^{mn} = K_2 K_\tau \dot{E}^{mn}  + K_1 E^{mn},\label{kate1}
\ee
with the boundary and initial conditions
\be 
\label{pBC} u^{mn}_i = U^{mn}_i(t)   +\Theta^{mn}(t) \varepsilon_{ij3} X_j \on \Gamma_{inc},\\
u^{mn}_i \quad \text{and} \quad \sigma_{ij}^{mn} n_j \quad \text{periodic} \on \Gamma_{box},\\
\int_{\Gamma_{inc}} \sigma^{mn}_{ij} {n}_j \, d\Gamma = 0,\qquad \int_{\Gamma_{inc}} \varepsilon_{ik3} \sigma^{mn}_{ij} X_k n_j \, d\Gamma = 0, \label{ohk}\\
\label{tool} \frac{\dd U^{mn}_i}{\dd X_j}  + \delta_{im} \delta_{jn} \delta(t-t') = 0,\qquad \frac{\dd \Theta^{mn}}{\dd X_j} = 0,\\
E^{mn}|_{t=0} = 0, \quad S^{mn}|_{t=0} = 0, \quad e_{ij}^{mn}|_{t=0} = 0, \quad s_{ij}^{mn}|_{t=0} = 0.
\ee} 

\paragraph{The cell function solutions} Equations (\ref{tool}) can be integrated to give 
\be \label{kate}
U^{mn}_i=-\delta_{im} \delta_{jn} X_j \delta(t-t') + \bar{U}^{mn}_i(t), \quad \Theta^{mn} = \Theta^{mn}(t),
\ee 
and $\bar{U}(t)$ remains to be determined from the force balance
integral constraint (\ref{ohk}). There are two problems which need to
be determined; we solve one for an excitation in the 11-direction and
another {with excitation} in the 12-direction, and will obtain
the remaining two solutions by rotation. These cell functions are
numerically evaluated in the frequency domain using the scheme
outlined in appendix \ref{implt}, where the transformed variables are
indicated with bar (see Fig.~\ref{MSP}). 

The solutions provided in Figs.~\ref{Cells}(a)--(c) {and denoted}  with hat correspond to the sum of the stress components shown in Figs.~\ref{MSC} and~\ref{MSP}, i.e., $\hat{s}_{ij}^{mn} = \tilde{s}_{ij}^{mn} + \bar{s}_{ij}^{mn}$ and $\hat{S}^{mn} = \tilde{S}^{mn} + \bar{S}^{mn}$.

\begin{figure}
\centering
\includegraphics[scale=0.5]{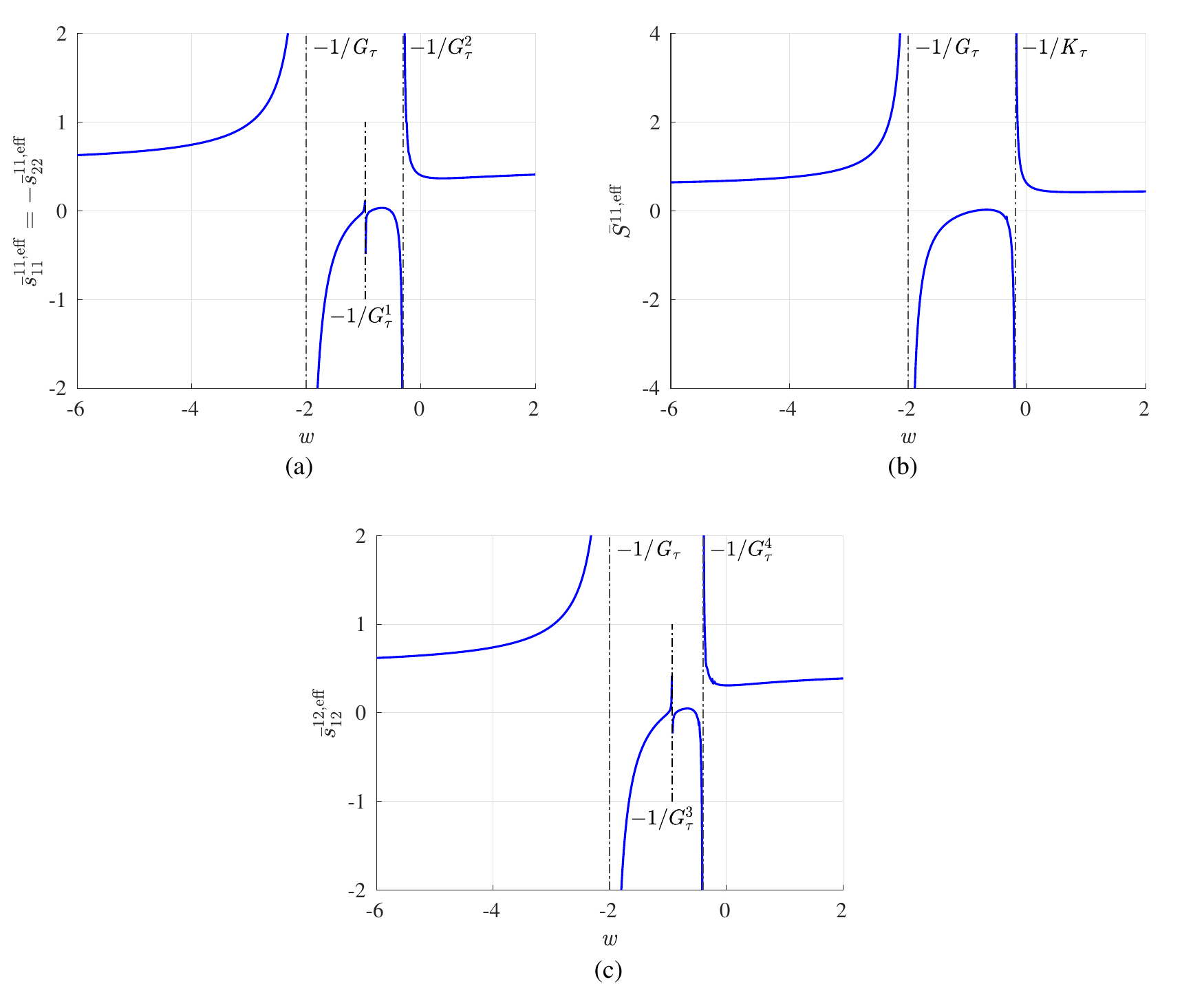}
\caption{The cell functions that characterise the response of the effective material due to deformation gradients in the particles {obtained as} solutions to equations~\eqref{kate0}--\eqref{kate} after application of the averaging operator~\eqref{averagingOp} and Laplace transformation.}
\label{MSP}
\end{figure}


\subsection{\label{binswe}{Response $\sigma^{\beta}_{ij}$ to binder swelling}}
The problem for the cell function which captures the deformations forced by the swelling of the binder, denoted with a superscript $\beta$, is 
\be
\frac{\partial \sigma_{ij}^{\beta}}{\partial X_j} =0, \quad \epsilon_{ij}^{\beta} = \frac{1}{2} \left( \frac{\partial u^{\beta}_i}{\partial X_j} + \frac{\dd u^{\beta}_j}{\dd X_i} \right),\label{binswe0}\\
S^{\beta} = \frac{1}{2} \sigma^{\beta}_{kk}, \quad E^{\beta} = \frac{1}{2} \epsilon^{\beta}_{kk},\\
s^{\beta}_{ij} = \sigma^{\beta}_{ij} - \delta_{ij} S^{\beta},\quad e^{\beta}_{ij} = \epsilon^{\beta}_{ij} - \delta_{ij} E^{\beta},\\
\label{kick1} {G_{\tau} \dot{s}^{\beta}_{ij} + s^{\beta}_{ij} = G_2 G_\tau \dot{e}^{\beta}_{ij} + G_1 e^{\beta}_{ij},} \quad {K_{\tau} \dot{S}^{\beta} + S^{\beta} = K_2 K_\tau \dot{E}^{\beta}  + K_1 E^{\beta} - \delta(t-t'),}
\ee
with the boundary and initial conditions
\be 
u^{\beta}_i = U^\beta_i(t)  +\Theta^{\beta}(t) \varepsilon_{ij3} X_j \on \Gamma_{inc},\\
u^{kl}_i \quad \text{and} \quad \sigma_{ij}^{kl} n_j \quad \text{periodic} \on \Gamma_{box},\\
\int_{\Gamma_{inc}} \sigma^{\beta}_{ij} {n}_j d\Gamma = 0, \qquad \int_{\Gamma_{inc}} \varepsilon_{ik3} \sigma^{\beta}_{ij} X_k n_j d\Gamma = 0,\\
\frac{\dd U^{\beta}_i}{\dd X_j} = 0, \qquad \frac{\dd \Theta^{\beta}}{\dd X_j} = 0,\\
E^{\beta}|_{t=0} = 0, \quad S^{\beta}|_{t=0} = 0, \quad e_{ij}^{\beta}|_{t=0} = 0, \quad s_{ij}^{\beta}|_{t=0} = 0. \label{binswe1}
\ee
Deformations are forced by the impulse at time $t=t'>0$ supplied via (\ref{kick1}b).

Applying a Laplace transform to (\ref{binswe0})--(\ref{binswe1})
results in a system of equations with {a similar form, except
  that all dependent variables are now functions of the transform variable and the terms involving time derivatives take an algebraic
  form. Equation (\ref{kick1}b) then} becomes
\be
(w K_\tau + 1) \hat{S} = (w K_\tau K_2 + K_1 ) \hat{E} - 1.
\ee 

\paragraph{The cell function solution} The solution to this cell problem can be determined analytically. Upon inverting the Laplace transform we find
\be
u^\beta_i = U^\beta_i = \Theta^\beta =0, \label{bssol0} \\
\label{bssol1}\epsilon^\beta_{ij} = 0,\quad e^\beta_{ij}=0, \quad E^\beta=0, \quad \sigma^\beta_{11} = \sigma^\beta_{22} = \frac{S^\beta}{2} \quad \sigma^\beta_{12}=0, \quad s^\beta_{ij} =0,\\ \label{bssol2} S^\beta = \begin{cases}
    0, & \for t<t', \\
    -\frac{1}{K_{\tau}} \exp \left( - \frac{1}{K_\tau} (t-t') \right) & \for t > t'.
  \end{cases}
\ee
The function $\hat{S}^{\beta,\text{eff}}$, which is what is needed to specify the effective material, is shown in Fig.~\ref{BS}.

\begin{figure}
\centering
\includegraphics[scale=0.5]{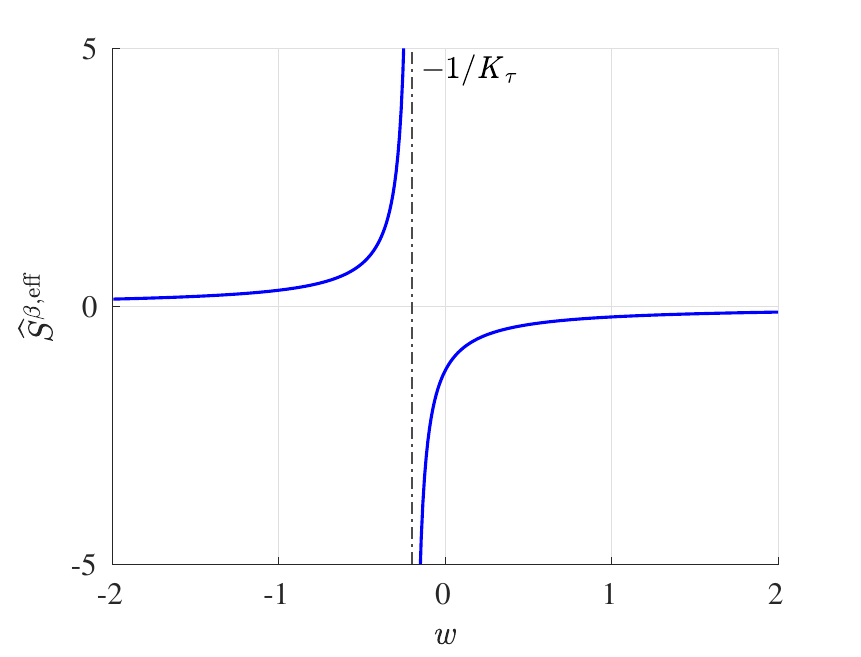}
\caption{The cell functions that characterise the response of the effective material due to binder swelling {obtained as} solution to equations~\eqref{binswe0}--\eqref{binswe1} after application of the averaging operator~\eqref{averagingOp} and Laplace transformation.}
\label{BS}
\end{figure}


\subsection{\label{actswe}{Response  $\sigma^{g}_{ij}$ to active particle swelling}}
The problem for the cell functions which capture the deformations forced by the swelling of the active particles, henceforth denoted with a superscript $g$, is
\be
\frac{\partial \sigma_{ij}^{g}}{\partial X_j} =0, \quad \epsilon_{ij}^{g} = \frac{1}{2} \left( \frac{\partial u^{g}_i}{\partial X_j} + \frac{\dd u^{g}_j}{\dd X_i} \right),\label{actswe0}\\
S^{g} = \frac{1}{2} \sigma^{g}_{kk}, \quad E^{g} = \frac{1}{2} \epsilon^{g}_{kk},\\
s^{g}_{ij} = \sigma^{g}_{ij} - \delta_{ij} S^{g},\quad e^{g}_{ij} = \epsilon^{g}_{ij} - \delta_{ij} E^{g},\\
{G_{\tau} \dot{s}^{g}_{ij} + s^{g}_{ij} = G_2 G_\tau \dot{e}^{g}_{ij} + G_1 e^{g}_{ij},} \quad {K_{\tau} \dot{S}^{g} + S^{g} = K_2 K_\tau \dot{E}^{g}  + G_1 E^{g},}\label{lala2}
\ee
with the boundary and initial conditions
\be 
\label{kick2} u^{g}_i = X_i \delta(t-t') + U^{g}_i(t)  +\Theta^{g}(t) \varepsilon_{ij3} X_j \on \Gamma_{inc},\\
u^{kl}_i \quad \text{and} \quad \sigma_{ij}^{kl} n_j \quad \text{periodic} \on \Gamma_{box},\\
\int_{\Gamma_{inc}} \sigma^{g}_{ij} {n}_j d\Gamma = 0, \qquad \int_{\Gamma_{inc}} \varepsilon_{ik3} \sigma^{g}_{ij} X_k n_j d\Gamma = 0,\\
\label{lala} \frac{\dd U^{g}_i}{\dd X_j} = 0,\qquad \frac{\dd \Theta^{g}}{\dd X_j} = 0,\\
E^{g}|_{t=0} = 0, \quad S^{g}|_{t=0} = 0, \quad e_{ij}^{g}|_{t=0} = 0, \quad s_{ij}^{g}|_{t=0} = 0.\label{actswe1}
\ee
Here, the forcing is provided by the boundary condition (\ref{kick2}a) at time $t=t'>0$.

\paragraph{The cell function solution} We can integrate (\ref{lala}) and then symmetry of the RVE implies that 
\be
U^g_i = \Theta^g = 0. 
\ee
The deformation, stress and strain fields need to be determined numerically and we make use of the finite element method to approximate solutions, as detailed in appendix \ref{implt}. A plot of the quantity $\hat{S}^{g,\text{eff}}$, which is required in the effective system of equations, is given in Fig.~\ref{APS}.

\begin{figure}
\centering
\includegraphics[scale=0.5]{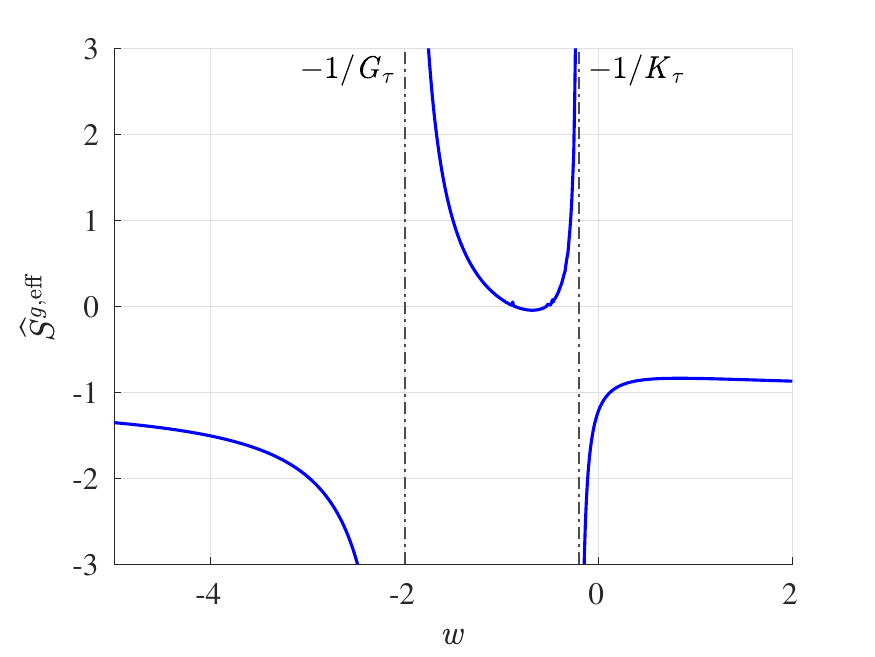}
\caption{The cell functions that characterise the response of the effective material due to swelling of the particle {obtained as} solution to equations~\eqref{actswe0}--\eqref{actswe1} after application of the averaging operator~\eqref{averagingOp} and Laplace transformation.}
\label{APS}
\end{figure}


\subsection{\label{mpg}{Response $p^q$ to macroscopic pressure gradients }}
{The problem for the cell functions which capture the deformations forced by macroscopic pressure gradients will henceforth be denoted with a superscript $q$, indicating a pressure gradient across the unit cell in the $q$-direction. The problem to be solved is
\be
\label{lapl} {\frac{\dd^2 p^q}{\dd X_i \dd X_i}}=0,
\ee
with the boundary conditions
\be 
 \left( \frac{\dd p^{q}}{\dd X_i} + e_q  \right) {n}_i = 0, \on \Gamma_{inc},\\ 
 p^q \quad \text{periodic on} \quad \Gamma_{box}, \label{lapl3}
\ee}
{where $e_q$ is the unit vector in the $q$th coordinate direction.}
Owing to the symmetry of our RVE we only need to solve for a single excitation, in the 1-direction, say, and the solution for an excitation in the 2-direction can be obtained by a rotation.

\paragraph{The cell function solutions} Once again, the solution is
determined numerically {as described in appendix \ref{implt}} and
is shown in Fig.~\ref{pressure}. This cell function solution is used
to specify the effective permeability using the definition
\eqref{kappaeff} and results in the curve shown in Fig.~\ref{kappa}.

\begin{figure}
\centering
\includegraphics[scale=0.7]{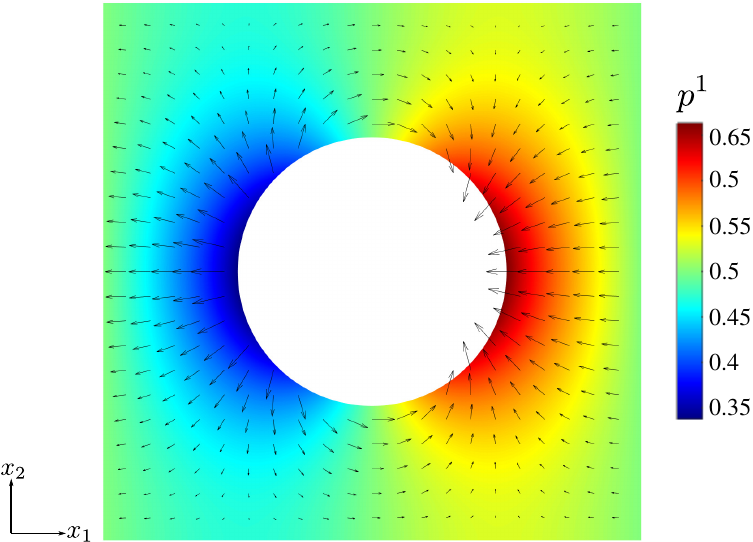}
\hspace{-2cm}
\caption{Pressure distribution $p^1$ and the velocity vector field
  $v^1 = (dp^1/dx_1, dp^1/dx_2)$ {corresponding to macroscopic pressure gradients with $q = 1$.}}
\label{pressure}
\end{figure}

\begin{figure}
\centering
\includegraphics[scale=0.5]{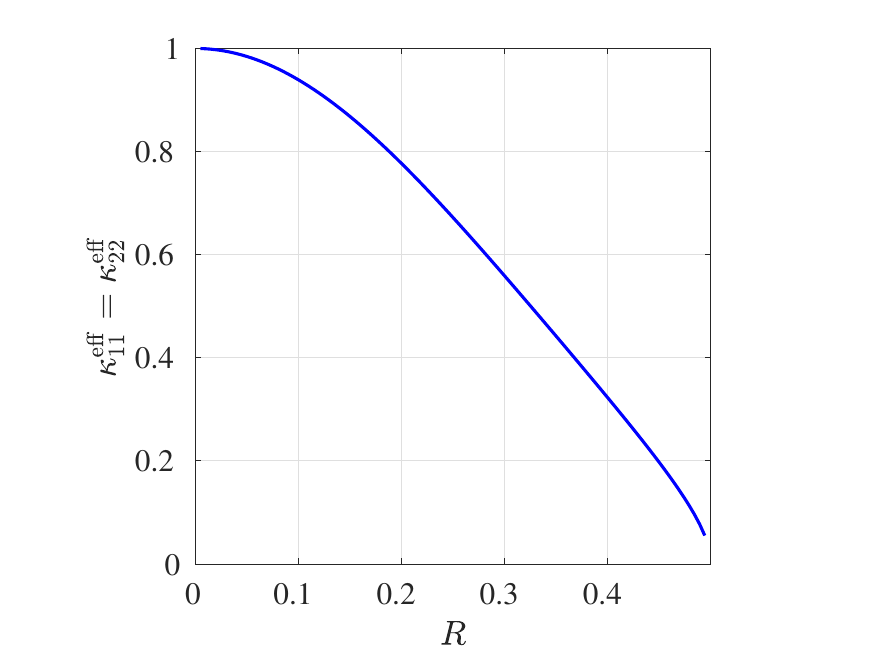}
\caption{{Dependence of the cell function characterising} the response of the effective permeability {on the particle radius $R$ obtained via} \eqref{kappaeff}.}
\label{kappa}
\end{figure}

\section{Numerical methods}\label{implt}
The finite element method is employed throughout this work via the
open-source software \verb{FreeFEM++{ \cite{Hecht}. Rather than
    tackling the time-dependent problem directly, we carry out all
    computations in the (complex) frequency domain by  applying a Laplace transform, which replaces the viscoelastic
    problem with one of elasticity. {However, since we are
      interesting in complex values of $w$,  the  additional
      terms have  complex-valued coefficients}. On the application of a Laplace transform (indicated with a hat) to (\ref{benny})--(\ref{noemmys}) and summing the result we arrive at 
\be \label{stress}
\hat{\sigma}_{ij}(w,\hat{{\bf u}}) = \frac{G_2 G_\tau w + G_1}{G_\tau {w} + 1} \hat{e}_{ij}(\hat{{\bf u}}) + \frac{K_2 K_\tau w + K_1}{K_\tau {w} + 1} \delta_{ij}\hat{E}(\hat{{\bf u}}),
\ee
{where $w$ is the transform variable, and $\hat{{\bf u}}$ represents
  the displacement field in the Laplace domain}; \eqref{stress} is identical in form to the familiar constitutive equation of isotropic elasticity. This removes the need to carry out time stepping thereby reducing computational complexity. A detailed explanation of how to solve elasticity problems, along with worked examples, is presented in the \verb{FreeFEM++{ user's guide.

The geometry for calculation of the cell functions is shown in Fig.~\ref{RVEmodel}, where $\alpha = 0.25$. The mesh parameters $k_1$, $k_2$, and $N$, and element types are presented in Table~\ref{Tab3}. The geometry for solution of the full model is shown in Fig.~\ref{model} and numerical parameters in Table~\ref{Tab2}. Computations were executed using the parallel version of {\tt MUMPS}. Convergence and accuracy tests were carried out and the numerical solutions were found to be reliable to at least 3 significant figures throughout.

\begin{figure}
\centering
\includegraphics[scale=0.35]{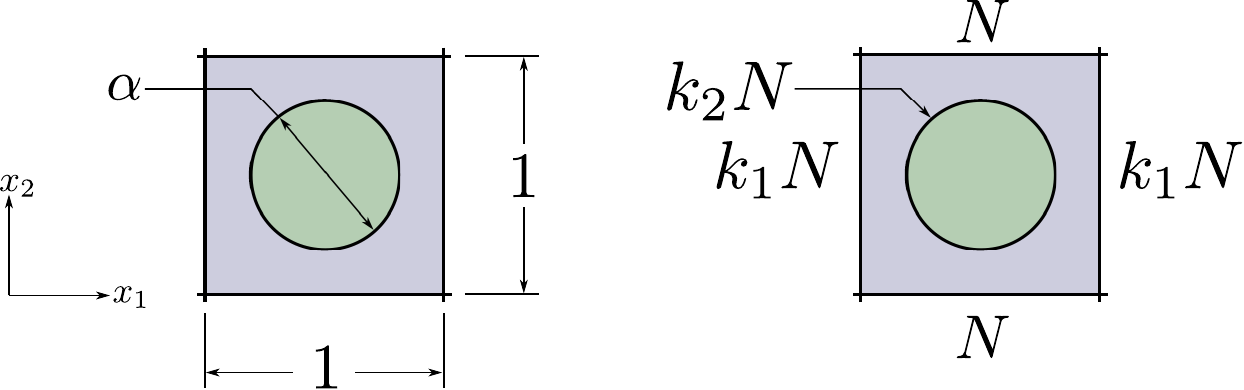}
\caption{RVE for cell function calculations.}
\label{RVEmodel}
\end{figure}

\begin{table}
\centering
        \begin{tabular}{c|ccccc}
        \toprule[1.2pt]
         \textit{Cell function} & \textit{Element} & $k_1$ & $k_2$ & $N$ & $N_T$ \\
        \midrule\midrule
		(i) & \verb{P1{ & $1$ & $1$ & $100$ & $12,760$ \\
		(ii) & \verb{P2{ & $1$ & $1$ & $450$ & $256,406$ \\
		(iii) & \verb{P1{ & $1$ & $1$ & $100$ & $12,760$ \\
		(iv)  & \verb{P2{ & $1$ & $1$ & $100$ & $32,294$ \\
 		\bottomrule[1.2pt]
        \end{tabular}
\caption{ Parameters for mesh generator used in \texttt{FreeFEM++}, {where $N$ is the number of elements in each segment, and} $N_T$ is the total number of elements used in the simulations. }\label{Tab3}
\end{table}

\begin{figure}
\centering
\includegraphics[scale=0.35]{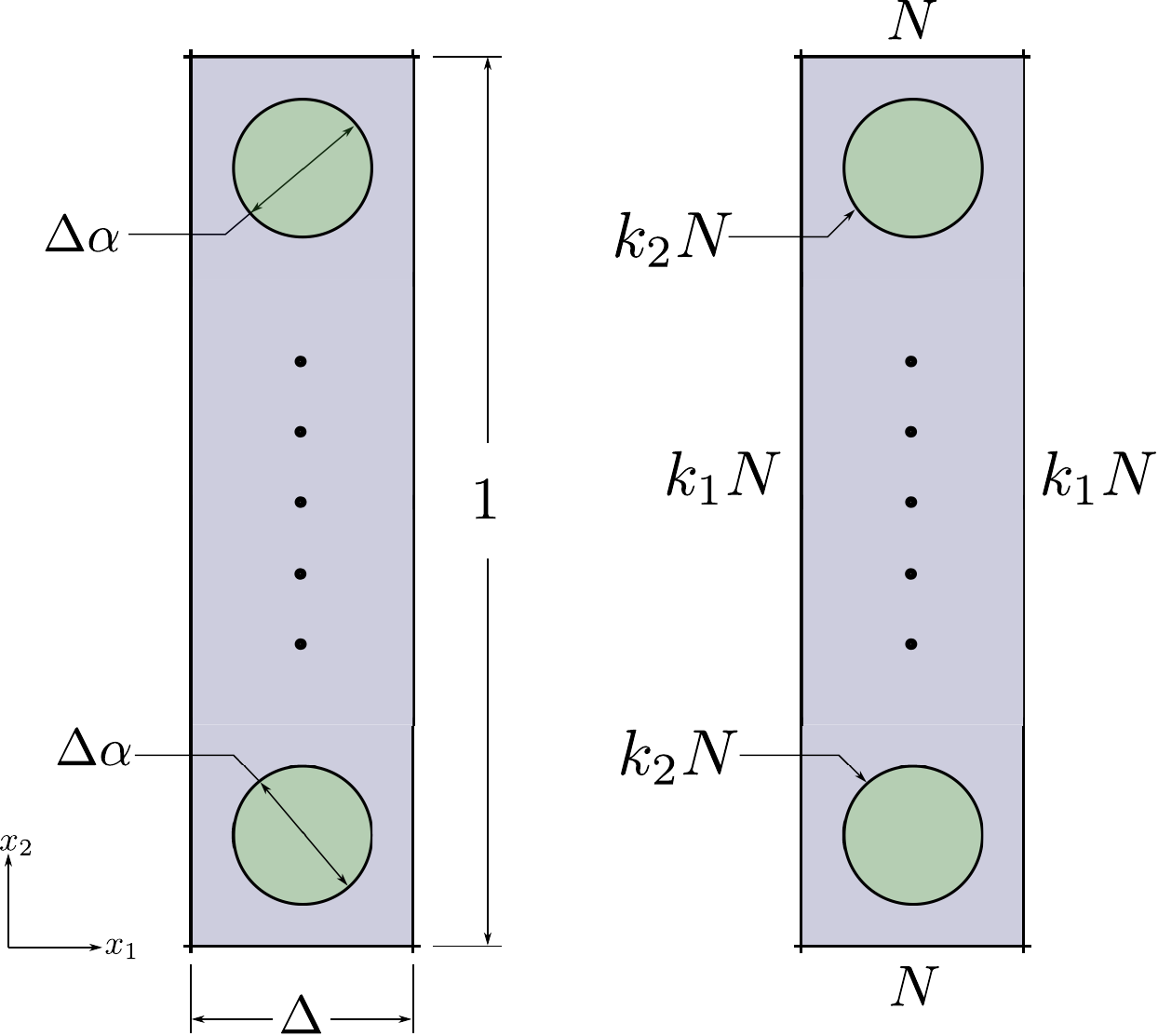}
\caption{Geometry for validation cases.}
\label{model}
\end{figure}

\begin{table}
\centering
        \begin{tabular}{c|ccccc}
        \toprule[1.2pt]
        $\Delta$ & $1$ & $1/2$ & $1/4$ & $1/8$ & $1/16$ \\
        \midrule\midrule
		$N$ & $100$ & $20$ & $14$ & $14$ & $7$ \\
		$k_1$ & $1$ & $17$ & $21$ & $17$ & $28$ \\
		$k_2$ & $2\pi$ & $2\pi$ & $2\pi$ & $2\pi$ & $3.6\pi$\\
		$N_T$ & $54,756$ & $34,322$ & $24,874$ & $24,628$ & $24,060$\\
 		\bottomrule[1.2pt]
        \end{tabular}
\caption{ Parameters for the mesh generator used in \texttt{FreeFEM++}.}\label{Tab2}
\end{table}

\end{document}